\documentclass[aps, prl, twocolumn, superscriptaddress, english, nobibnotes]{revtex4-2}

\usepackage[english]{babel}

\usepackage[dvipsnames]{xcolor}

\usepackage{amsmath, amssymb}
\usepackage{mathtools}
\usepackage{amsthm}
\usepackage{amscd}
\usepackage{graphicx}
\usepackage[colorlinks=true]{hyperref}
\hypersetup{pdfencoding=auto,colorlinks=true, allcolors = blue}

\usepackage[utf8]{inputenc}
\usepackage{dsfont}
\usepackage[T1]{fontenc}
\usepackage{braket}
\usepackage{float}

\usepackage{enumerate}
\usepackage{textcomp}
\usepackage{soul}
\usepackage[normalem]{ulem}

\newcommand{\setZ}{\mathds{Z}}
\newcommand{\setN}{\mathds{N}}

\newcommand{\vdag}{{\phantom\dag}}

\DeclareMathOperator{\Tr}{Tr}

\DeclareMathOperator{\unit}{\mathds{1}}

\def\be{\begin{equation}} 
\def\ee{\end{equation}}

\def\ba{\begin{align}} 
\def\ea{\end{align}}

\def\f{\varphi} 
\def\wh{\widehat} 
\def\de{\partial} 

\newcommand{\opN}{\widehat{N}}
\newcommand{\Ham}{\widehat{H}}
\newcommand{\On}{\widehat{\mathcal{O}}_{n}}

\newcommand{\Up}{{a}}
\newcommand{\Dn}{{b}}

\newcommand{\pos}{j}
\newcommand{\spec}{\alpha}

\usepackage{tikz}

\usetikzlibrary{calc}
\usetikzlibrary{decorations.pathreplacing,calligraphy}
\usetikzlibrary{fit}
\tikzset{
    aten/.style={
        draw,
        fill=MidnightBlue!20,
        inner sep = 0,
        minimum width = 0.6cm,
        minimum height = 0.6cm
}}

\tikzset{
    oins/.style={
        circle,
        draw,
        fill=Mulberry!20,
        inner sep = 0,
        minimum width = 0.6cm,
        minimum height = 0.6cm
}}
\tikzset{
    tmat/.style={
        draw,
        fill=PineGreen!20,
        inner sep = 0,
        minimum width = 0.6cm,
        minimum height = 0.6cm
}}

\tikzset{
    ncbar angle/.initial=90,
    ncbar/.style={
        to path=(\tikztostart)
        -- ($(\tikztostart)!#1!\pgfkeysvalueof{/tikz/ncbar angle}:(\tikztotarget)$)
        -- ($(\tikztotarget)!($(\tikztostart)!#1!\pgfkeysvalueof{/tikz/ncbar angle}:(\tikztotarget)$)!\pgfkeysvalueof{/tikz/ncbar angle}:(\tikztostart)$)
        -- (\tikztotarget)
    },
    ncbar/.default=0.5cm,
}

\graphicspath{{./}{./Figure/}}

\begin{document}

\title{Quantum simulation of the tricritical Ising model in tunable Josephson junction ladders}

\author{Lorenzo Maffi}
\thanks{These authors contributed equally to this work.}
\affiliation{Center for Quantum Devices and Niels Bohr International Academy, Niels Bohr Institute, University of Copenhagen, DK--2100 Copenhagen, Denmark}
\affiliation{Dipartimento di Fisica e Astronomia “G. Galilei”,
Universit\`a degli Studi di Padova, I-35131 Padova, Italy
3}
\affiliation{Istituto Nazionale di Fisica Nucleare (INFN), Sezione di Padova, I-35131 Padova, Italy}

\author{Niklas Tausendpfund}
\thanks{These authors contributed equally to this work.}
\affiliation{Forschungszentrum J\"{u}lich GmbH, Institute of Quantum Control,
Peter Gr\"{u}nberg Institut (PGI-8), 52425 J\"{u}lich, Germany}
\affiliation{Institute for Theoretical Physics, University of Cologne, D-50937 K\"{o}ln, Germany}

\author{Matteo Rizzi}
\affiliation{Forschungszentrum J\"{u}lich GmbH, Institute of Quantum Control,
Peter Gr\"{u}nberg Institut (PGI-8), 52425 J\"{u}lich, Germany}
\affiliation{Institute for Theoretical Physics, University of Cologne, D-50937 K\"{o}ln, Germany}

\author{Michele Burrello}
\affiliation{Center for Quantum Devices and Niels Bohr International Academy, Niels Bohr Institute, University of Copenhagen, DK--2100 Copenhagen, Denmark}

\begin{abstract}
Modern hybrid superconductor-semiconductor Josephson junction arrays are a promising platform for analog quantum simulations. Their controllable and non-sinusoidal energy/phase relation opens the path to implement nontrivial interactions and study the emergence of exotic quantum phase transitions. Here, we propose the analysis of an array of hybrid Josephson junctions defining a 2-leg ladder geometry for the quantum simulation of the tricritical Ising phase transition. This transition provides the paradigmatic example of minimal conformal models beyond Ising criticality and its excitations are intimately related to Fibonacci non-Abelian anyons and topological order in two dimensions. We study this superconducting system and its thermodynamic phases based on bosonization and matrix-product-states techniques. Its effective continuous description in terms of a three-frequency sine-Gordon quantum field theory suggests the presence of the targeted tricritical point and the numerical simulations confirm this picture. Our results indicate which experimental observables can be adopted in realistic devices to probe the physics and the phase transitions of the model. Additionally, our proposal provides a useful one-dimensional building block to design exotic topological order in two-dimensional scalable Josephson junction arrays.
\end{abstract}

\maketitle

The rapid advances in the fabrication of superconducting/semiconducting heterostructures \cite{krogstrup2015,shabani2016} allow for the realization of Josephson junction arrays (JJAs) with unprecedented tunability of their physical parameters \cite{bottcher2018,bottcher2022,Bottcher2022b}.
State-of-the-art electron beam lithography and etching techniques enable the realization of superconducting (SC) arrays with exquisite geometrical precision and scalability.
Epitaxial growth consents to create pristine interfaces between a semiconducting substrate and SC islands, thus providing the possibility of controlling these setups through voltage gates. 
These fabrication developments are flanked by remarkable advances in measurement techniques which include microwave spectroscopy to study the strongly correlated systems emerging in Josephson junction chains \cite{Bell2018,manucharyan2019,higginbotham2022} and transport measurements to investigate the intricate thermodynamic properties of these systems \cite{cedergren2017,bottcher2018,bottcher2022,Bottcher2022b,higginbotham2022}.
Such progresses brought JJAs right back into the arena of analog quantum simulation platforms, where they started their journey decades ago.
The simultaneous tunability of the junction transparencies \cite{shabani2016,Kjaergaard2017,Casparis_NatNanoTech2018,Ciaccia2023,Banszerus2024} and magnetic fluxes opens indeed the path to tailor models of interest, among which quantum field theories (QFTs) and integrable models \cite{Bell2018,Roy2019,Saleur2021,Roy2023.1}.
In particular, 
the experimental achievement of multicritical points,
with peculiar conformal field theories (CFTs) associated with them \cite{difrancesco}, becomes within reach~\cite{Roy2023.2}.

In this work, we formulate a blueprint for the quantum simulation of the tricritical Ising (TCI) CFT in a tunable Josephson junction ladder.
The reasons for interest in this model are multiple.
It constitutes the simplest example of CFT beyond the Ising model, and its particle content includes excitations that share the same fusion properties of Fibonacci non-Abelian anyons.
Successfully implementing this model will open the way to engineer exotic topological order in 2D arrays in the spirit of the wire constructions of Refs. \cite{Alicea2014,Alicea2015,Kane2018,oreg-franz20}. 
Moreover, the TCI model stands as a strong potential candidate to observe the emergence of supersymmetry \cite{Friedan1984,Rahmani2015,o'brien2018}.
Notably, to our knowledge, no experimental realization of a quantum TCI phase transition in 1D has ever been observed, nor have its critical exponents been measured.

Indeed, the quantum simulations of CFTs beyond the Ising universality class face both experimental and theoretical challenges: 
the most recent theoretical proposals rely on advanced constructions based on Majorana modes \cite{Rahmani2015,Rahmani2015.prb,zhu2016,o'brien2018,ebisu2019,oreg-franz20}, extended Hubbard models with staggering potentials \cite{essler16,Ejima2018b} or nontrivial mappings between microscopic lattice operators and the field content of the CFTs \cite{Mong2014}. In this context, the main mechanism to achieve a TCI point is to consider platforms like Rydberg atom systems \cite{slagle2021,slagle2022} and ultracold atoms in tilted optical superlattices \cite{Buyskikh2019} that are described by discrete models with a continuous Ising phase transition turning into a first-order phase transition (FOPT) at the tricritical point.

JJAs offer a direct way to implement the scaling limit of interacting bosonic QFTs \cite{Saleur2021,Roy2023.2}. In the following we present a ladder system that embodies a three-frequency sine-Gordon model and can be tuned to naturally flow towards the TCI point at low energy. The chosen ladder geometry offers an alternative construction compared to previous works on SC chains \cite{Roy2023.1,Roy2023.2} (see also the ladder construction in Ref.~\cite{Bell2018}), and opens a path towards 2D devices with exotic properties \cite{slagle2022}. 
To achieve our goal, we utilize a blend of analytical techniques, including mean field analysis and bosonization \cite{Giamarchi2003}, complemented by numerical results based on variational uniform matrix product states (VUMPS) \cite{Haegeman2011, Haegeman2016, Stauber2018}.

\begin{figure}
\centering
\centering
\includegraphics[width=1\columnwidth]{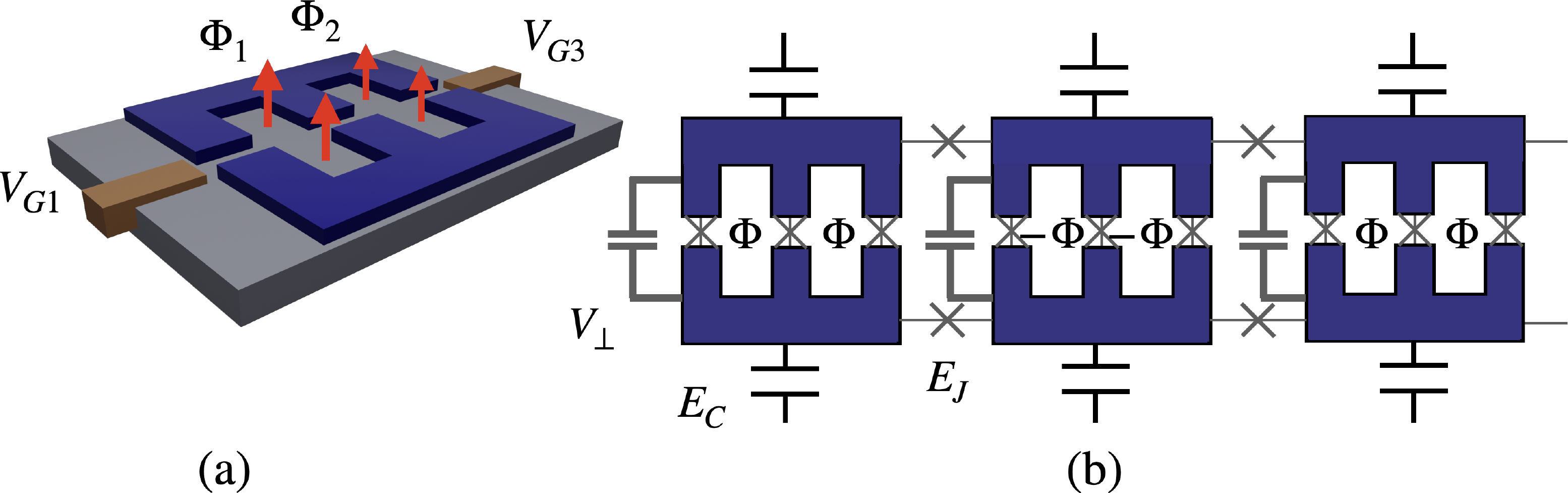}
\caption{(a) Two E-shaped SC islands are connected through three parallel junctions. An out-of-plane magnetic field (red arrows) dictates the Aharonov-Bohm phases $\Phi_1$ and $\Phi_2$ along the two loops.
The external junctions are controlled by electrostatic gates at potential $V_{G1}$, $V_{G3}$ which vary the carrier density in the surrounding semiconductor. This triple JJ element allows us to control the potential \eqref{V3cos} at each rung of the ladder geometry (b). 
The fluxes of the triple JJ elements are staggered along the ladder \cite{Supplemental}.  Mutual rung capacitances and the island self-capacitances determine the electrostatic interactions $V_\perp$ and $E_C$.} 
\label{fig1}
\end{figure}

\paragraph{The triple Josephson junction.-}
The building block of our 1D construction consists of two E-shaped SC islands facing each other and grown on a semiconducting substrate [Fig.~\ref{fig1}(a)]. Schematically, we model this element as three parallel Josephson junctions (JJs) \cite{Supplemental} where Andreev bound states induced in the semiconductor mediate the Cooper pair tunneling \cite{beenakker1991,marcus2018}. For simplicity, we assume that each junction is defined by a single transport channel with transparency $T_p \in [0,1]$ ($p=1,2,3$) and energy/phase relation \cite{beenakker1991}: 
\begin{equation}
\label{beenakker}
\mathcal{E}_J^{(p)}\left(\varphi\right)=-\Delta\sqrt{1-T_p\sin^2{\left(\f/2\right)}}\,,
\end{equation}
See also Refs. \cite{Bozkurt2023,Banszerus2024} for alternative realizations. In Eq. \eqref{beenakker}, $\f$ is the phase difference between the two islands and $\Delta$ is the SC gap induced by proximity in the semiconducting substrate. 
High-transparencies $T_p$ lead to coherent tunneling events of multiple Cooper pairs \cite{Heikkila2002} corresponding to higher harmonics contribution, $\cos(n\f)$ with $n>1$, to the dispersion \eqref{beenakker}. 
In the  triple JJ geometry, the amplitudes of such events can be tuned by inserting two magnetic fluxes in the resulting loops [Fig.~\ref{fig1}(a)] \cite{Supplemental}.

We set $\Phi_1=\Phi_2=\Phi$ and identical transparencies ($T_1=T_3$) for the external junctions, controlled using electrostatic gates [Fig.~\ref{fig1}(a)].
With these constraints, the exchange of the SC islands, $\varphi\to -\varphi$, corresponds to the required $\setZ_2$-symmetry for the multicritical Ising physics, which is reflected in the odd current/phase relation of the triple JJ.
Multiple channels in the junctions or unequal plaquette areas may explicitly break this symmetry~\cite{Supplemental}, hindering the observation of critical features whenever the corresponding energy gaps are larger than the experimentally achievable energy resolution due to the finite size $L$ and the temperature. 
In the symmetric setup, the total Josephson potential can be expanded as
\begin{equation} \label{V3cos}
    V_J\left(\f\right)=\sum_{n\in\setN}\mu_{n}({\bf X})\cos{\left(n\f\right)}.
\end{equation}
The Fourier coefficients $\mu_n$ \cite{Supplemental} depend on the values of the external parameters ${\bf X}=\left(T_1\cos{\left(\Phi\right)},\,T_1\sin{\left(\Phi\right)},\,T_2\right)$ which span a solid cylinder.

\begin{figure}
\centering
\includegraphics[width=\columnwidth]{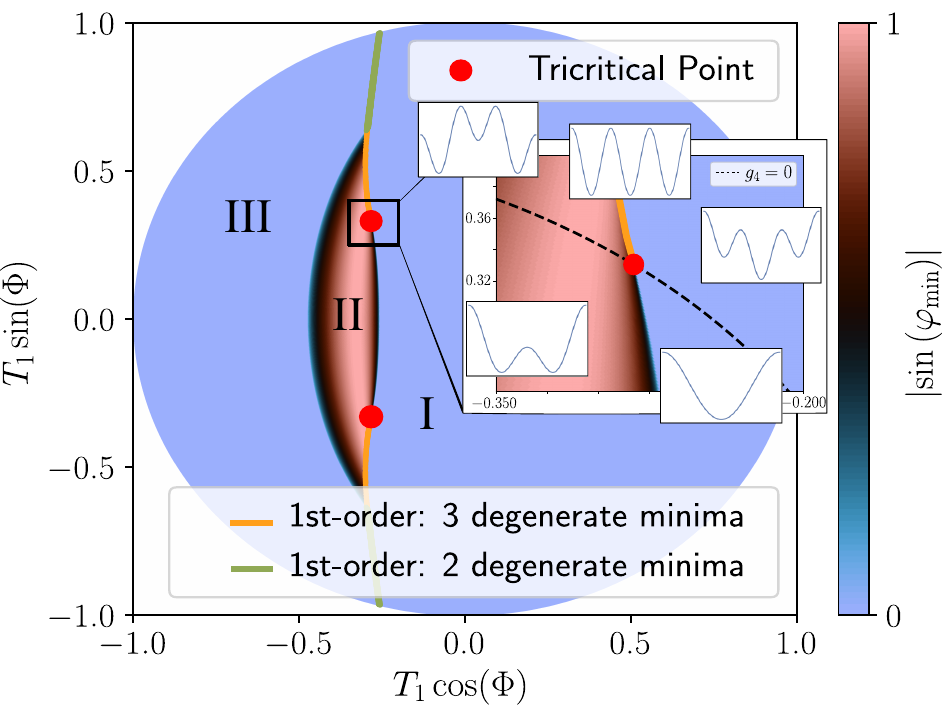}
\caption{Given $\f_{\rm min}$ the global minimum of $V_J$ in Eq. \eqref{V3cos}, we depict $\left|\sin{\left(\varphi_{\rm min}\right)}\right|$ in the parameter space at $T_2=0.6$. 
Regions I and III correspond to $\setZ_2$-symmetric configurations with $\f_{\rm min}=0,\pi$ respectively. Region II presents two degenerate minima. Inset: the transition between region I and II can be either discontinuous with three degenerate minima (yellow line) or continuous with the merging of the two minima in $\varphi_{\rm min}=0$. The red dot labels a tricritical point where a three-well potential $V_J=g_2 \varphi^2+g_4\varphi^4+\varphi^6$ approximates Eq. \eqref{V3cos}.
The dashed line corresponds to $g_4=0$.
}
\label{fig2}
\end{figure}

We will use many copies of this triple JJ to build a 1D ladder geometry, thus promoting the phase difference $\f$ to a position-dependent field. In light of this, a preliminary mean-field analysis allows us to qualitatively understand the onset of a TCI point by investigating the potential $V_J\left(\f\right)$ as a function of ${\bf X}$. In a semiclassical picture, a tricritical point arises when three potential minima merge \cite{zamolodchikov1986,Lassig1991,lepori2008}. In the landscape defined by $V_J(\f)$ with $\f\in\left(-\pi,\pi\right]$, for any $T_2$, there exists a point $\left(T_1,\Phi\right)_c$ where this merging occurs and $V_J(\f)$ is approximated by a $\f^6$ local potential, see Fig.~\ref{fig2}. 
This suggests the first connection to the TCI model and its Ginzburg-Landau (GL) formulation \cite{zamolodchikov1986,Lassig1991,lepori2008}.

\paragraph{1D model.-}

We design a 1D quantum simulator to achieve a TCI point by arranging a set of identical triple JJs with potential $V_J$ in parallel, as depicted in Fig.~\ref{fig1}(b), to implement a multiple-frequency sine-Gordon model at low energies.
The Hamiltonian of the JJ ladder is:
\begin{equation}\label{Ham}
    \begin{split}
\Ham=\sum_{\pos=0}^{L-1}\Biggl[  
    \sum_{\spec=\Up,\Dn}&\!\left( 
                E_C \opN_{\spec,\pos}^2 
              - E_J\cos{\left(\hat{\f}_{\spec,\pos+1} -\hat{\f}_{\spec,\pos} \right)} 
          \right)
        \Biggr. \\ 
    {}&+  \Biggl. V_{\perp}\,\opN_{\Up,\pos}\opN_{\Dn,\pos} 
    +V_J\left(\hat{\f}_{\Up,\pos}-\hat{\f}_{\Dn,\pos}\right)\Biggr],
\end{split}
\end{equation}
where $\hat{\f}_{\spec,\pos}$ represents the phase operator of the $j$-th island on the leg $\spec\in\left\{\Up,\Dn\right\}$.
Along the legs, the SC islands are connected through JJs in a standard sinusoidal regime with Josephson energy $E_J$. This energy scale can vary from $E_J\simeq h \, 50\;$GHz \cite{Casparis_NatNanoTech2018} down to $E_J=0$ for completely depleted junctions. The dynamics of the SC phases in Eq. \eqref{Ham} is dictated by charging effects, described by the charge operators $\opN_{\spec,\pos}$, canonically conjugated to the SC phases, $[\opN_{\spec,\pos},e^{i\hat{\f}_{\spec,\pos}}]=-e^{i\hat{\f}_{\spec,\pos}}$. We consider in particular an on-site electrostatic repulsion $E_C$ and a rung repulsive interaction $V_\perp$. 

To obtain the rung potentials $V_J$ in Eq. \eqref{Ham}, the pattern of magnetic fluxes in the system must be carefully considered: a uniform magnetic field breaks time-reversal invariance driving the system into Meissner chiral phases \cite{Orignac2001,Petrescu2013,Piraud2015,Petrescu2015,Greschner2015,Greschner2016,Haller2020} and does not fulfill the $\mathbb{Z}_2$-symmetry on each rung. We consider instead staggered fluxes alternating at each triple JJ [Fig. \ref{fig1}(b)]. This choice yields the local effective potential \eqref{V3cos} and avoids additional fluxes between subsequent rungs \cite{Supplemental}.

The aimed multi-frequency sine-Gordon model emerges when the rung potentials $V_J$ and the Josephson energy $E_J$ dominate over the charging effects $E_C$ and $V_\perp$. 
In this Josephson-dominated regime, the system lies away from Mott insulating phases \cite{Fazio2001rev,Goldstein2013,Petrescu2015} and phase localization due to charge disorder \cite{Giamarchi1987,Orignac1998,Crepin2011} is strongly irrelevant. The effects of disorder in the potential $V_J$ are discussed in \cite{Supplemental}.
In the continuum limit, the low-energy physics of the Cooper pairs can be described through bosonization \cite{Giamarchi2003} by introducing dual fields $(\hat{\theta}_{\spec}(x),\hat{\f}_\spec(x))$ for each leg $\spec$, with $\left[\hat\theta_\alpha(y),\hat\f_\beta(x) \right] = -i \pi \delta_{\alpha \beta} \Theta\left(y-x\right)$. $\opN_{\spec,\pos}/a\approx -\de_x\hat{\theta}_{\spec}(x)/\pi$ represents the charge of the island $j=x/a$ and $a$ the lattice spacing. 

By defining the customary charge $c$ and spin $s$ sectors, $\hat{\f}_{c/s}(x)=\left(\hat{\f}_{\Up}(x)\pm \hat{\f}_{\Dn}(x)\right)/\sqrt{2}$, the Hamiltonian \eqref{Ham} is approximated by \cite{Supplemental}:
\begin{multline} 
\label{SG}
    \Ham = \sum_{q=c,s} u_{q}\int \dfrac{dx}{2\pi}\left[K_q\left(\de_x\hat{\f}_{q}\right)^2+\dfrac{1}{K_q}\left(\de_x\hat{\theta}_q\right)^2\right]\\+\int \frac{dx}{a}\,\sum_{n=1}^3\mu_n\cos{\left(\sqrt{2}n\hat{\f}_s\right)}.
\end{multline}
Eq. \eqref{SG} describes the two branches of the model as Luttinger liquids (LLs), with Luttinger parameters $K_{c/s}\approx\pi\sqrt{E_J/\left(2E_C\pm V_{\perp}\right)}$ \cite{Orignac2001,Petrescu2015}. The rung potential $V_J$ affects only the spin branch and yields the targeted multiple sine-Gordon interactions. 
The three potential terms in Eq. \eqref{SG} must be relevant in the renormalization group sense and induce
order in the phase $\hat{\f}_s$, driving the spin sector away from the LL phase. This sets the constraint $K_s > 9/4$, which, indeed, is fulfilled for sufficiently large Josephson energies, when the semiclassical description is most accurate.  
Higher harmonics in Eq. \eqref{V3cos}, instead, are neglected as less relevant and characterized by smaller amplitudes \cite{Supplemental}. 

The interplay of the three sine-Gordon terms $\mu_n$ yields nontrivial phase transitions \cite{Delfino1998,Toth2004,Roy2023.2} between the low-energy massive phases of the spin sector. 
In particular, an Ising critical line meets a FOPT in a tricritical point characterized by the TCI CFT with central charge $c=7/10$ \cite{Toth2004,Roy2023.2}.

\begin{figure*}[htb]
    \centering
    \includegraphics[width =\textwidth]{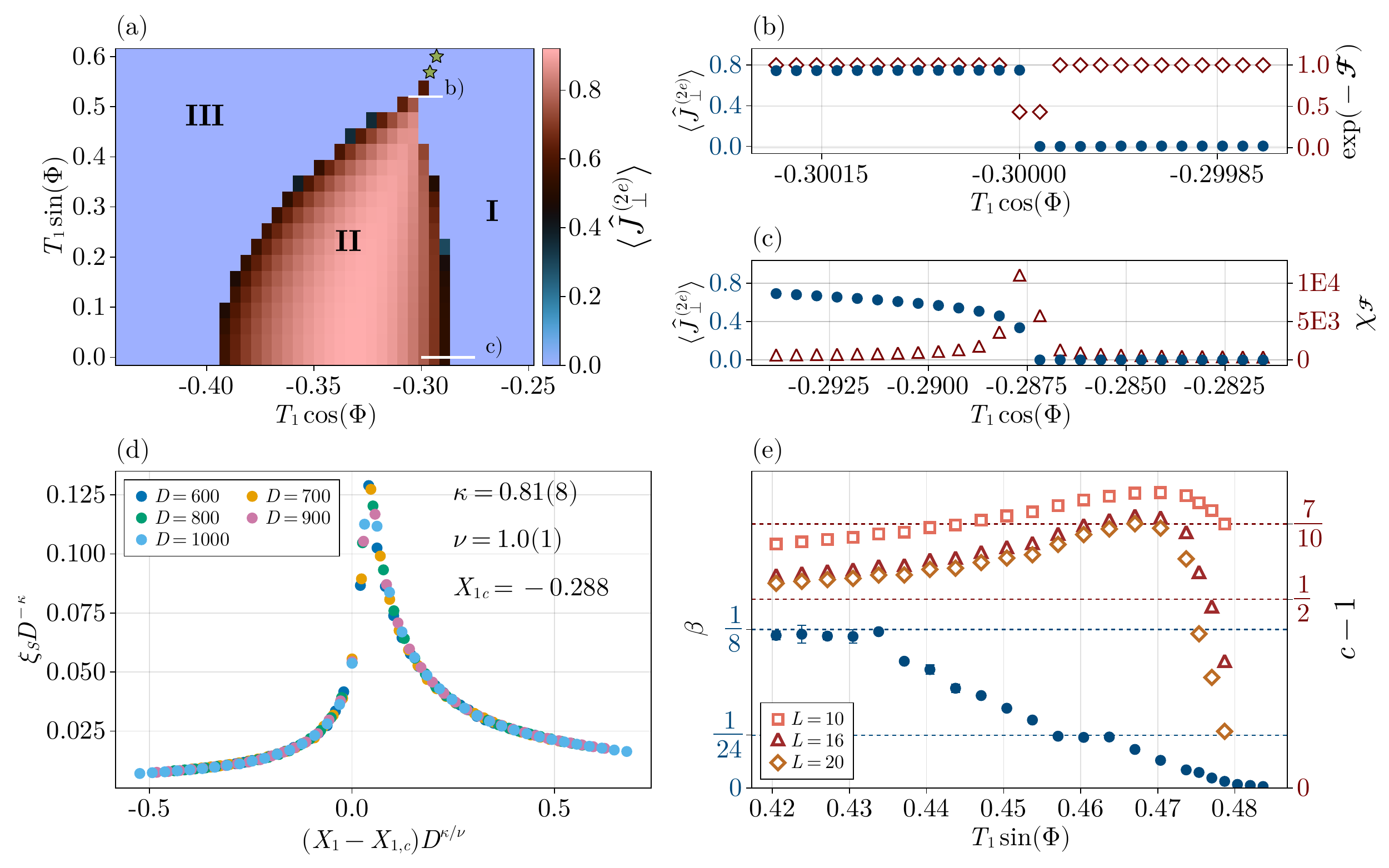}
    
    \caption{(a): Expectation value of the order parameter $\hat{J}^{(2e)}_\perp$ at $T_2=0.6$. Green stars mark a discontinuity of the log-fidelity per site [Eq.~\eqref{Fidelity}] denoting the FOPT between phases I and III, consistently with the mean-field picture. (b):  FOPT discontinuity of $\exp{\left(-\mathcal{F}\right)}$ and $\langle\hat{J}^{(2e)}_\perp \rangle$ between phases II and I at $X_2=0.52$ [cut b) in panel (a)].  (c): singular behavior of the fidelity susceptibility $\chi_\mathcal{F}$ and order parameter along the cut c) at $X_2=0$, both indicating a second-order phase transition. 
    (d): collapse of the correlation length $\xi_s$ at $X_2=0$ for five values of the bond dimension $D$ by employing a finite-entanglement scaling \cite{Tagliacozzo2008,Supplemental}.
    (e): critical exponent $\beta$ obtained by fitting $\langle\hat{J}^{(2e)}_\perp \rangle$ as a function of $X_1$ for $0.42<X_2<0.49$ and bond dimension $D=600$ (blue dots). Two plateaux appear close to the Ising $(\beta_{\rm IS}=1/8)$ and TCI $(\beta_{\rm TCI}=1/24)$ predictions. The central charge (empty symbols), derived from finite-size DMRG simulations \cite{Supplemental}, increases from $c\simeq1+1/2$ to $c\simeq1+7/10$ before dropping to $c\simeq1$.}
    \label{fig3}
\end{figure*}

\paragraph{Observables and results.-} We study the phase diagram of our model by using the variational uniform matrix product
state ansatz (VUMPS), \cite{Haegeman2011, Haegeman2016, Stauber2018}, to find the ground state of the Hamiltonian~\eqref{Ham} in the thermodynamic limit.
The VUMPS is based on a two-site elementary cell representing two SC islands on the same rung. The local Hilbert space is constructed from the charge basis defined by $\opN_{\spec=\Up/\Dn,\pos}$. For numerical purposes, we truncate its basis by introducing a cutoff, $\left| N_{\spec,\pos}\right|< N_{\rm max}$, with $N_{\rm max}\ge 6$ \cite{Supplemental}. 

We set $E_C/E_J=0.4$ and $V_\perp/E_J=0.65$, corresponding to $K_s\approx 8$. This favours the clean emergence of the transition lines as the interactions are strongly relevant, yielding sizeable energy gaps in the spin sector. 
The Fourier components $\mu_n$ in Eq. \eqref{V3cos} are determined from Eq. \eqref{beenakker} with a SC gap $\Delta/E_J=50$ and $T_2=0.6$, consistent with Fig.~\ref{fig2}.

We identify the phases of the model with labels I, II and III as in Fig. \ref{fig2}, and, to distinguish them, we employ the local order operator $\hat{J}^{(2e)}_{\perp}(x)=\sin{\left(\sqrt{2}\hat{\f}_s(x)\right)}$ representing the single-particle contribution to the rung current.
In the VUMPS simulations, the symmetry-broken phase II is signaled by a finite $\braket{\hat{J}^{(2e)}_\perp}$ [Fig.~\ref{fig3}(a)], and it aligns with the mean-field predictions in Fig.~\ref{fig2}. The symmetric phases I and III broaden away from the semiclassical limit due to the dominant scaling behavior of the first-harmonic interaction.  
The order parameter allows us to investigate the boundary between the disordered phase I and the ordered phase II: a neat jump in $\braket{\hat{J}^{(2e)}_\perp}$ marks a FOPT for $X_2=T_1\sin{\left(\Phi\right)} \gtrsim 0.475$ [Fig.~\ref{fig3}(b)], while a continuous change in the region $\left|X_2\right| \lesssim 0.475$ indicates the onset of a second-order transition, as exemplified for $X_2=0$ in Fig.~\ref{fig3}(c). 

This picture is confirmed by the analysis of the ground state fidelities \cite{Zanardi2006,Cozzini2007,Rams2011,Rossini2018}. Given the abrupt change of the ground state $\ket{\psi\left({\bf X}\right)}$ across the FOPT, the average log-fidelity per site \cite{Rams2011}
\begin{equation}
    \mathcal{F}\left(\bf X, \delta\right)=-\lim_{N\to\infty}\dfrac{1}{N}\log\left(\braket{\psi({\bf X} -\delta)|\psi({\bf X}+\delta)}\right),
    \label{Fidelity}
\end{equation}
displays a clean discontinuity [Fig.~\ref{fig3}(b)], at fixed $\delta$.
 On the other hand, across the lower cut the fidelity susceptibility $\chi_\mathcal{F}=\mathcal{F}/\delta^2$ shows a more gradual singular behaviour and exhibits the typical peak of a second-order phase transition in Fig.~\ref{fig3}(c).

The universal collapse of the spin correlation length $\xi_s$ according to finite entanglement scaling ansatz \cite{Tagliacozzo2008,Supplemental} confirms that the continuous phase transition lies within the Ising universality class, see Fig.~\ref{fig3}(d): for $X_2=0$, we located the critical point $X_{1c}$ and extrapolated the infinite bond dimension estimate of the critical exponent $\nu=1.0(1)$, matching the CFT prediction $\nu_{\rm IS}=1$.
Additionally, our analysis reveals the scaling of the effective magnetization \cite{Supplemental} $\braket{\hat{J}^{(2e)}_\perp}~\sim~\left|X_1-X_{1c}\right|^{\beta}$, with the critical exponent $\beta$ compatible with the Ising value $\beta_{\rm IS}=1/8$ for $|X_2|<0.43$ [Fig. \ref{fig3}(e)].

The latter confirms also the onset of the TCI point joining the Ising phase transition and the FOPT: by increasing $X_2$ above $0.43$, $\beta$ decreases and, at $X_2 \sim 0.46$, it exhibits a plateau close to the expected TCI value $\beta_{\rm TCI}=1/24$ [Fig \ref{fig3}(e)]. Further increasing $X_2$ results in a vanishing $\beta$, as expected for a FOPT. The error bars in Fig.~\ref{fig3}(e) do not account for finite entanglement effects, accentuated by the massless LL in the charge sector with $c=1$ throughout the entire phase diagram. Despite this, we observe a good convergence in scaling features away from the critical point. 

Finally, along the transition line for $X_2>0.42$, finite-size density-matrix renormalization group (DMRG) simulations reveal in Fig.~\ref{fig3}(e) the non-monotonic behavior of the central charge $c$ \cite{Ejima2018,essler16}, consistently with the presence of the TCI CFT ($c-1=7/10$) amid the Ising regime ($c-1=1/2$) and the FOPT ($c-1=0$). Finite size effects yield large central charge estimates as expected and shift the tricritical point to larger $X_2$ relative to the $\beta=\beta_{\rm TCI}$.

\paragraph{Experimental observables.-} Transport features can be used to explore the phase diagram of the model. Indeed, the thermal conductance across 1D systems at criticality is proportional to the central charge $c$ of the related CFT at low temperature $T$ \cite{bernard2016,gawedzki2018}: $G_Q= \frac{\pi k_B^2 T c}{6\hbar}$. 
In our model, symmetric and symmetry-broken phases exhibit $c=1$ due to the charge sector, while along the transition line, the additional contribution of the spin sector yields the behaviour shown in Fig.~\ref{fig3}(e).

In thermal transport experiments \cite{partanen2016,gubaydullin2022}, heat currents will be dominated by the QFT collective modes for temperatures considerably below the SC gap ($\sim 2$K for Al). Finite size and temperature will affect the profile of the heat conductance as a function of the system parameters.
Nevertheless, a non-monotonic behavior of $G_Q$ across the second-order phase transition line and in proximity of the TCI point would provide strong evidence of the emergence of the related CFTs.

Furthermore, as the rung currents exhibit quasi long-range order at the phase transitions, the power spectrum of their noise provides a probe to detect the critical lines and measure the scaling dimension of the order parameter.
Additionally, microwave spectroscopy of JJAs \cite{Bell2018,manucharyan2019,higginbotham2022} allows for the study of the excitation spectra of the system and can be used to verify the predictions  of the TCI CFT spectra \cite{Feiguin2007,lepori2008,Vidal2018,Cubero2021,lencses2022}

\paragraph{Conclusions.-}
We designed a JJ ladder to realize a quantum simulator for the tricritical Ising CFT. Our construction is based on the properties of hybrid semiconducting-superconducting JJs and their non-sinusoidal energy/phase relation. In particular, we engineered a triple JJ that allows us to tune the higher harmonics and we adopted them to realize the physics of a multi-frequency sine-Gordon QFT \cite{Toth2004}.

We used bosonization and tensor-networks simulations to investigate this JJA. Our analysis showed the presence of an ordered phase and highlighted the existence of a critical Ising plane connected to a first-order transition along a tricritical Ising line within a three-parameter space.

Our construction does not require the introduction of strong and fine-tuned interactions and relies on the adjustments of parameters that can be controlled in hybrid state-of-the-art platforms. 

Our study poses the basis for further explorations of the connection between nontrivial interacting CFTs and hybrid JJ systems characterized by high harmonics terms. The ladder we devised, in particular, provides a tool to engineer systems with exotic topological order in two-dimensional setups: an array of these tricritical systems opens the way to realize Fibonacci topological superconductors \cite{oreg-franz20,Kane2018} with universal non-Abelian anyons.

\begin{acknowledgments}
\paragraph{Acknowledgements.-} 
We thank L. Banszerus, A. Cappelli, C. Marcus, G. Mussardo, C. Schrade and S. Vaitiekenas for fruitful discussions.
We acknowledge support from the Deutsche Forschungsgemeinschaft (DFG) project Grant No. 277101999 within the CRC network TR 183 (subprojects B01 and C01).
L.M. and M.B. are supported by the Villum Foundation (Research Grant No. 25310).
N.T. and M.R. are further supported by the DFG under Germany's Excellence Strategy - Cluster
of Excellence Matter and Light for Quantum Computing (ML4Q) EXC 2004\slash 1 – 390534769. The authors gratefully acknowledge the Gauss Centre for Supercomputing
e.V. (www.gauss-centre.eu) for funding this project by providing computing time through the John von Neumann Institute for Computing (NIC) on the GCS Supercomputer JUWELS at the J\"{u}lich Supercomputing Centre (JSC) (Grant NeTeNeSyQuMa) and the FZ J\"{u}lich for JURECA (institute project PGI-8) \cite{JURECA2021}. Data and Code are available at \cite{maffi_2023_10225786}.
\end{acknowledgments}

\newpage

\appendix

\setcounter{equation}{0}

\renewcommand{\theequation}{S\arabic{equation}}
\renewcommand{\thefigure}{S\arabic{figure}}

\onecolumngrid

\section*{Supplemental materials}

\section{Triple Josephson junction element}

\subsection{Higher harmonics expansion}

In this section, we briefly analyze the decomposition of the energy-phase relation of the triple JJ into harmonic terms $\mu_{n}$ that we introduced in Eq. \eqref{V3cos} 
of the main text. 
Assuming that each semiconducting/superconducting junction is described by a single quantum channel, the potential of triple JJ element
\be
    \label{potential}
    V_J\left(\varphi\right)=-\Delta\left( 
                \sqrt{1-T_1 \sin^2\left(\frac{\varphi - \Phi_1}{2}\right)}
               +\sqrt{1-T_2 \sin^2\left(\frac{\varphi\vphantom{\Phi}}{2}\right)}
               +\sqrt{1-T_3 \sin^2\left(\frac{\varphi + \Phi_2}{2}\right)}
            \right),
\ee
can be expanded as $V_J=\sum_n \mu_{n}\cos{\left(n\f\right)}$, where $\f$ is the SC phase difference of the two islands and $\Delta$ the superconducting gap induced in the semiconducting layer of the hybrid system. To maintain  the reflection symmetry 
$\varphi\to-\varphi$, we impose $\Phi_1=\Phi_2=\Phi$ and $T_1=T_3$.
The full expression of $\mu_{n}$ involves the elliptic integrals
\be
    \label{ellip}
    \mu_{n}=\int_{-\pi}^{\pi}\dfrac{d\f}{\pi}\,V_J\left(\f\right)\cos{\left(n\f\right)},
\ee
which do not have an elementary analytical solution. However, for small transparencies $T_i\ll 1$, we can approximate them as follows:
\be
    \label{Ejn}
    \begin{split}
        \mu_{1}/\Delta &=- \dfrac{1}{512}\left( T_2\left(128+32T_2+15T_2^2\right)+2T_1\left(128+32T_1+15T_1^2\right)\cos{\Phi}\right)+O\left(T_i^4\right)\\
        \mu_{2}/\Delta&=\dfrac{1}{256}\left(T_2^2\left(4+3T_2\right)+2T_1^2\left(4+3T_1\right)\cos{2\Phi}\right)+O\left(T_i^4\right)\\
        \mu_{3}/\Delta&=-\dfrac{1}{512}\left(T_2^3+2T_1^3\cos{3\Phi}\right)+O\left(T_i^4\right)\\
        \mu_{4}/\Delta&=O\left(T_i^4\right).
    \end{split}
\ee
In this limit, it is evident that the potential $V_J$ is mostly determined by the first harmonic term $\cos{\f}$ with $\mu_{1}<0$, as long as the magnetic flux is such that $\cos{\Phi}>0$. Numerical evaluation of the integrals \eqref{ellip} shows that this is true also in the large transparencies limit. 

\begin{figure}[h]
    \centering
    \includegraphics[width = \textwidth]{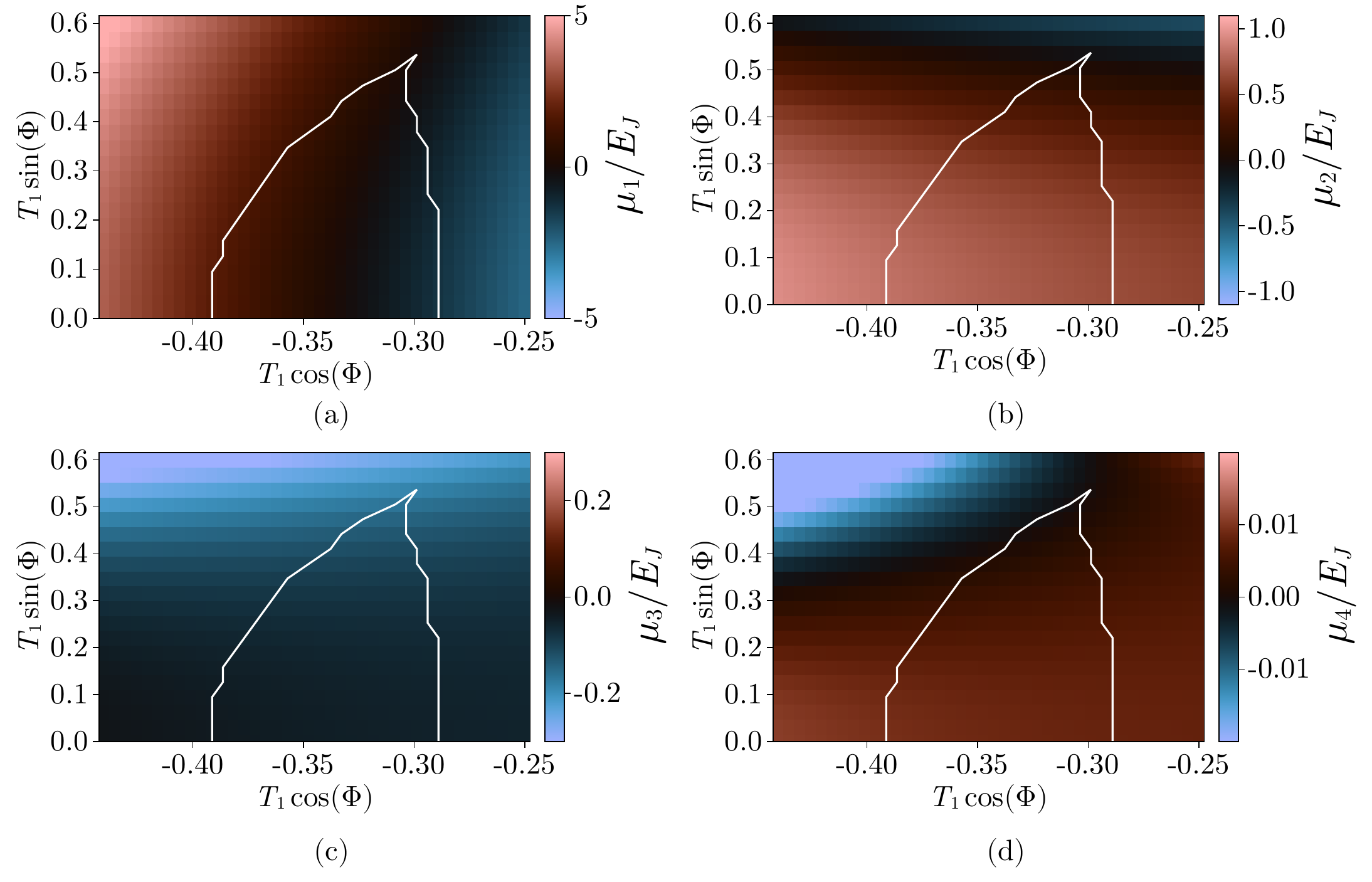}
    \caption{The amplitudes of the first four harmonics $\mu_{n}$ with $n=0,1,\ldots,4$ as a function of the triple JJ parameters. The white lines mark the boundary of the symmetry-broken regime. We set $T_2=0.6$ and the SC gap $\Delta$ induced by proximity in the semiconductors is fixed at $\Delta=50$ in units of $E_J$ and does not influence the ratio between the $\mu$ coefficients \eqref{ellip}.}
    \label{SMfig1}
\end{figure}

The situation is different if we consider fluxes such that $\cos{\Phi}<0$. In particular, one can fine-tune the external parameters to make $\mu_{1}$ vanish. Moreover, for $\Phi=2\pi/3$ and $T_1=T_2$ both $\mu_{1}$ and $\mu_{2}$ vanish as a consequence of destructive interference of tunneling events of one and two Cooper pairs through the three junctions. In this case only triplet of Cooper pairs can jump between the two SC islands with amplitude $|\mu_{3}|$. One can also check that, in the considered geometry, the contribution $\mu_{4}$ is always at least one order of magnitude smaller than the other terms as showed in Fig.~\ref{SMfig1}. 
Therefore, given the ability of controlling both the transparencies of the hybrid junctions through external gates and the magnetic flux piercing the two loops, we can tune independently the ratios between the first three harmonics amplitudes in Eq. \eqref{potential}. In particular, the results discussed in the main text require that only the transparencies of the external junctions, $T_1$ and $T_3$, need to be tuned, whereas $T_2$ does not qualitatively affect the appearance of the tricritical Ising point. This constitutes an advantage for experimental realizations since we envision that the external junctions can more easily be controlled via electrostatic gates.

Importantly, our approximations hold when each junction is sufficiently shorter than the (diffusive) coherence length of the superconducting regions induced in the semiconductor, allowing coherent tunneling process. This is achieved in \cite{Ciaccia2023} with a length of 150 nm. The width of the junction, instead, mostly affects the amount of active quantum channels in the junction: the limit of single-channel junction has been experimentally investigated in hybrid nanowire devices, with widths of about 100 nm \cite{marcus2018,hart19}.

\subsection{Multichannel case} 

In the case of several transport channels in each of the junctions, the Josephson energy-phase relation is given by the sum of the related contributions:
\begin{equation}
    \mathcal{E}_{J}^{(p)}= -\sum_{i=1}^{M_p} \Delta\sqrt{1-T_p^{(i)}\sin^2{\left(\phi/2\right)}},
\end{equation}
where $T_p^{(i)}$ represents the transparency of the $i$th channel in the JJ $p$, and $M_p$ is the number of channels in the junction. For disordered multichannel junctions, these transport coefficients $T_p^{(i)}$ follow a bimodal distribution \cite{Beenakker1997}, with a few high-transparency channels resulting in a nonsinusoidal current response. A complete generalization of our results to the multichannel case goes beyond the scope of this supplemental section. However, a qualitative analysis of its effects is needed. In particular, one essential feature of our triple JJs element is the symmetry between the two external junctions.

Experimental results for wide junctions (with width $W\simeq 2-3\,{\rm\mu m}$) in gate-tunable device showed that the nonsinusoidal effects are overall well-approximated by one JJ with $M^*$ high-transparency channels with the same average $T^*$, such that the current phase relation reads \cite{Nichele2020,Ciaccia2023}
\begin{equation}
\label{multichannel}
    I\left(\f\right)= \dfrac{e \Delta M^*T^*}{\hbar}\dfrac{\sin{\left(\f\right)}}{\sqrt{1-T^* \sin^2{\left(\f/2\right)}}}.
\end{equation}
Therefore, the nonlinear function in Eq. \eqref{beenakker} 
in the main text well approximates the energy-phase relation also in the multichannel case.
Equation \eqref{multichannel} represents a phenomenological approximation that effectively described the behavior of past experimental platforms \cite{Ciaccia2023}, but it does not capture comprehensively the multichannel case.

In such approximation, one can assume that the external voltage gate $V_G$ affects only the number of channels $M^*$ and not the average transparency $T^*$, which mildly varies among the junctions \cite{Ciaccia2023}. In this case, the symmetry between the external JJs is lifted by the weak finite difference between the two average transparencies $T^*_1- T^*_3\neq 0$, which is almost independent of the voltage gates $V_{G1}$ and $V_{G3}$. However, tuning the number of open channels $M^*_1$ and $M^*_3$ via the voltage gates provides a way to mitigate this explicit symmetry breaking. Finally, potential asymmetries in the magnetic fluxes cause a splitting in energy of the minima of the potential $V_J$ which is linear in $\Phi_1-\Phi_3$. However, this effect can also be used to mitigate the asymmetry caused by the mismatch of the transparencies $T_1^*\neq T_3^*$ and restore the degeneracy of the minima of $V_J$.

Alternatively, as briefly mentioned in the main text, the non-sinusoidal current/phase relation can effectively be obtained by substituting each of the junctions with two sinusoidal multichannel JJs in series \cite{Bozkurt2023,Banszerus2024}. For the external links, the effective transmissions $T_{p,\rm eff}$ with $p=1,3$ will depend on the critical currents flowing through such JJs and indeed can be tuned by external electrostatic gates.

\section{Ladder: further details}

\subsection{Staggered magnetic fluxes}

Interacting bosons on a ladder with uniform magnetic fields exhibit are characterized by the onset of several chiral many-body phases, including the Meissner phase. For our purposes the onset of the Meissner effect may be detrimental, because it breaks the emergent Lorentz invariance in the QFT and may compete with the phases and critical points discussed in the main text. 

Additionally, to obtain a quantum simulation of the three-frequency sine-Gordon model, each rung triple JJ must be characterizes by the same $V_J$. This condition is, in the general case, fulfilled only by staggered patterns of magnetic fluxes.

\begin{figure}[t]
    \centering
    \includegraphics[width=1\textwidth]{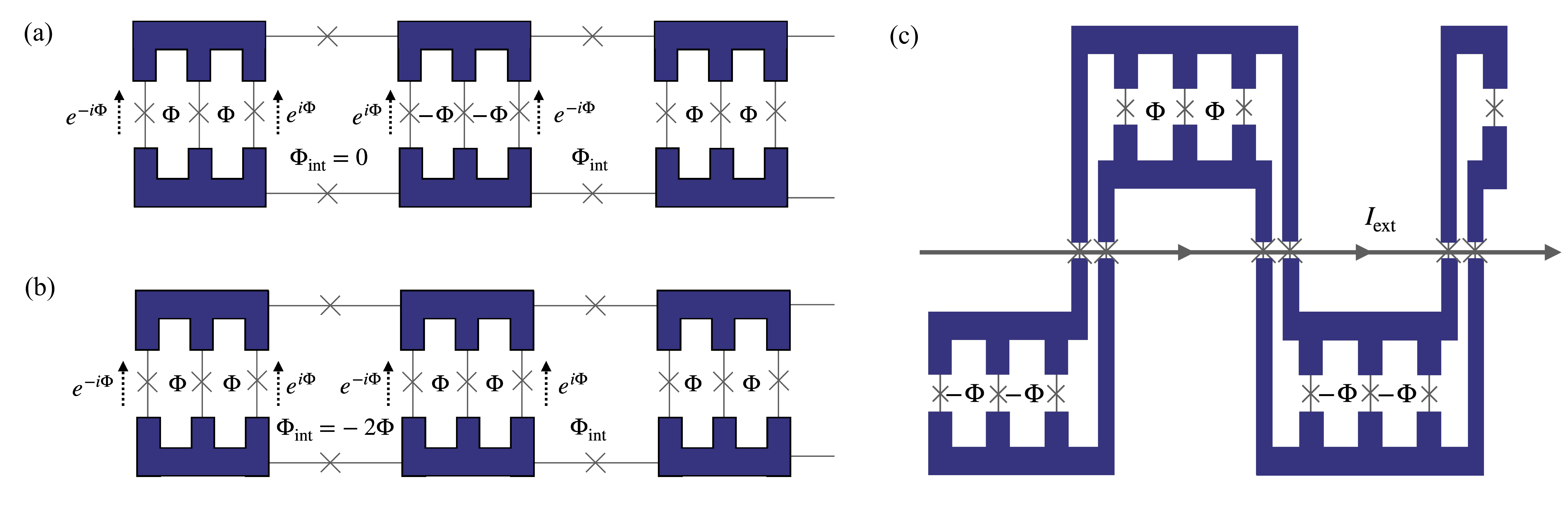}
   \caption{(a) and (b) illustrate the two configurations of nonuniform magnetic fluxes essential for preserving time-reversal invariance in the effective ladder description. In configuration (a), the fluxes are staggered within consecutive triple Josephson junction (JJ) elements, while in (b), the $\Phi_{\rm int}=-2\Phi$ condition is implemented in the plaquettes of the effective ladder. Panel (c) shows the physical realization of configuration (a) achieved through a snake geometry and the insertion of a line with a tunable external current $I_{\rm ext}$.}
    \label{SMfig2}
\end{figure}

We present two viable flux configurations which are schematically represented in Fig.~\ref{SMfig2}(a) and (b). The solution (a) relies on the parity property of the local potential $V_J$ under $\Phi\to-\Phi$ and enables the engineering of a ladder geometry where the magnetic flux between two subsequent rungs, thus the related Aharonov-Bohm phase $\Phi_{\rm int}$, vanishes. This preserves time-reversal invariance in the effective QFT. However, this approach leads to the experimental challenge of controlling nonuniform magnetic fields along the ladder.

A convenient construction to realize the configuration (a) in experimental devices is depicted in Fig.~\ref{SMfig2}(c). To stagger the magnetic fluxes within two subsequent triple JJ elements, we design the ladder in a 'snake' configuration and control the magnetic field by introducing a current $I_{\rm ext}$ through the line schematically represented in Fig.~\ref{SMfig2}. Alternatively, a local control of multiple fluxes can be achieved with the techniques adopted by modern quantum processors based on transmon qubits \cite{Google1}.

An alternative flux configuration, Fig.~\ref{SMfig2}(b) results in the same potentials $V_J$ on each rung and relies on compensating the magnetic fluxes of the triple JJs with opposite fluxes in the ladder plaquettes, thus setting $\Phi_{\rm int}=-2\Phi$ between each rung. The possibility of introducing additional integer fluxes in each loop, thus replacing $\Phi_{\rm int}\to \Phi_{\rm int}+2\pi$ may also offer an alternative to implement the configuration (b) with uniform magnetic fluxes. To tune the system at the tricritical point in this scenario, however, it is required to known a priori the parameter $T_2$ of the ladder: the critical flux of the trijunctions depends indeed on $T_2$; therefore, its knowledge is necessary to designing superconducting circuits with a correct ratio between the areas of the loops inside the trijunctions and the areas of the loops between the ladder rungs to obtain the desired tunneling phases at constant magnetic field.

\subsection{Disorder}

In the hybrid solid-state devices we consider, disorder is limited by the accurate epitaxial growth and lithographic techniques employed for their fabrication. Nevertheless, a certain amount of disorder is unavoidable due to the typical etching procedures adopted to define the Josephson junctions and it may prevent the emergence of the targeted many-body phases. In our physical device we envision two potential sources of disorder: (a) disordered-induced charges on the superconducting islands, and (b) disorder in the junction transmissions.
Given the large values of the Luttinger parameter $K_s$, we expect to be protected against the charge disorder (a) that results in irrelevant operators in the low-energy limit of the model.
On the other hand, the disorder (b) translates into a disordered local potential $V_J$, Eq. \eqref{V3cos} in the main text, and requires a more careful analysis.

In our proposal, we assume that the transmission $T_2$ of the central junction cannot be controlled, making it the primary source of this kind of disorder. A random distribution of $T_2$ maintains the $\mathbb{Z}_2$-symmetry of the ladder, while inducing random variations in the parameters $\mu_n$ in Eq. \eqref{V3cos} of the main text. When assuming Gaussian random disorder, we can give a rough estimate of the threshold over which disorder dominates over the features studied in our model by comparing their standard deviations with the typical gaps observed in the system.

In particular, when considering the gapped symmetry-broken phase, the impact of disorder can be estimated in the following way. Given a certain amount of disorder $\delta T_2/T_2$, we compare the energy scale $\Delta \delta T_2/T_2$ with the mass of the solitons interpolating between the two minima of $V_J$ in the related field theory, which provides a good approximation of the spin gap $\Delta_s$. In the semiclassical approach, the local potential $V_J$ approximately assumes the typical double well form $g_2\varphi^2+g_4 \varphi^4$, within the ordered phase II ($g_2<0$). By following standard calculations \cite{mussardobook}, we determine the soliton mass to be
\begin{equation}
    M_s=\dfrac{2\sqrt{2}}{3}\dfrac{\left|g_2\right|^{3/2}}{g_4}\sqrt{E_J},
\end{equation}
where we accounted for the Luttinger kinematics renormalization in spin sector (see the next subsection). The stability of the ordered phase hinges on whether the energy scale of the disorder in $T_2$ remains below $M_s$ and $\Delta_s$.
By considering the input values of our simulations, we derive that a $10\%$ disorder in $T_2$ constitutes the threshold over which the ordered phase is obscured, possibly leading to glassy physics phenomena. Similar results are obtained by comparing the disorder energy scale with the numerical gaps derived from the transfer matrix eigenvalues within the spin sector (see Sec. \ref{app:transfer} ).

Notably, however, such effects can be mitigated by increasing the Josephson energy scale $E_J$ along the legs of the ladder, thus $M_s \sim \Delta_s$. Larger values of $E_J$ decrease indeed the occurrence of phase slips in the 1D system.

In recent experimental systems with long JJ chains \cite{manucharyan2019}, characterized by more than 30000 JJs, the estimated disorder in the Josephson energies was below $10\%$. In this context, carefully engineered ladders with a smaller number of junctions fabricated to specifically observe the physics of the TCI should allow us to achieve the most favorable energy hierarchy for mitigating disorder effects and observe the many-body phases discussed in the main text.

A further useful fabrication aspect to emphasize in order to optimize the construction of the ladder device is the following: suitable amplitudes and large energy scales for the higher harmonics in the potential $V_J$ can be achieved by constructing triple junctions with a wider central junction with many low transmission channels, such that we approximate its energy-phase relation with the standard sinusoidal form $E_{J2}\cos{\varphi}$. By enlarging the size of the middle junctions, on one side we decrease the impact of geometric imperfections leading to disorder of the kind (b) and, on the other, we increase the energy gaps that characterize the gapped phases in our model, thus improving the resilience of the phase diagram against disorder.


Regarding the critical features of the ladder, they will remain clean below a characteristic disorder length that decreases with increasing disorder $\delta T_2$. If this disorder lengthscale becomes considerably smaller than the system size, however, unexpected critical scaling phenomena may emerge. A comprehensive understanding of disorder in conformal field theory (CFT) remains elusive, as does a systematic theoretical framework for its treatment.
Nevertheless, we can apply symmetry reasoning to our system and make use of the Harris criterion \cite{Harris1974} to provide qualitative insights.

According to the Harris criterion, a random quenched disorder that preserves the conformal symmetry becomes relevant only if it couples with a local operator of the CFT with scaling dimension $D<1$ \cite{Harris1974}; concerning disorder with a Gaussian distribution in general one-dimensional quantum systems, instead, the renormalization group analysis of Giamarchi and Schultz \cite{Giamarchi1988} shows that disorder is relevant if the related operator has dimension $D<3/2$. At the TCI point, the disorder in $T_2$ does not couple with the odd magnetizations $\sigma$ and $\sigma'$, which explicitly break the $\mathbb{Z}_2$-symmetry. This fact ensures the preservation of the ordered phase in the low-energy limit, preventing the system from losing long range order, analogously to what happens in the Ising CFT. Moreover, the disorder in $T_2$ couples with the less relevant thermal deformation $\epsilon$, with scaling dimension $1/5$. This implies that weak disorder introduces an additional lengthscale in the system, which diverges for clean systems and must be sufficiently large to observe criticality; the TCI features can be observed for distances below this disorder lengthscale, whereas observables extending over this length will present features typical of disordered and gapped systems.
To our knowledge, thermal disorder in TCI CFT has not been studied yet, in neither the classical nor the quantum case.

\subsection{Bosonization} \label{app:bos}

In this section, we will review the main steps of the connection between the lattice Hamiltonian in \eqref{Ham} 
in the main text and the three-frequency sine-Gordon quantum field theory. At low temperature $K_BT<\Delta_c$  each SC island of our lattice corresponds to a condensate of $N_c$ Cooper pairs with gap $\Delta_c$ and a well defined complex order parameter, the SC phase $\hat{\f}_{\spec,\pos}$. The residual charge around $N_c$ is represented by the operator $\opN_{\spec,\pos}$ dual to the SC phase. In the long wavelength limit, we can use an effective continuum description in terms of the Bose fields $\hat\theta_\alpha(x)$ and $\hat\f_\alpha(x)$ \cite{Giamarchi2003}, fulfilling commutation relations:
\begin{equation}
\left[\hat\theta_\alpha(y),\hat\f_\beta(x) \right] = -i \pi \delta_{\alpha \beta} \Theta\left(y-x\right)\,,
\end{equation}
where $\Theta$ indicates the Heaviside step function.
The weak interactions case $E_C,\;V_\perp,\;\ll E_J$ we considered allows us to neglect fast-oscillating contributions in the Cooper-pair density and write $\opN_{\spec,\pos}\approx -a\dfrac{\de_x\hat{\theta}_\spec(x)}{\pi}$, with $j=xa$.
In the harmonic approximation for the Josephson interaction along the legs, the low-energy lattice Hamiltonian can be written as
\begin{multline}
     \hat{H}= \sum_{\spec=\Up,\Dn}\left[\dfrac{E_J}{2}\int dx\;a\left(\de_x\hat{\f}_\spec\left(x\right)\right)^2+\dfrac{E_C a}{\pi^2} \int dx\;\left(\de_x\hat{\theta}_\spec\left(x\right)\right)^2\right]+ \frac{V_\perp a}{\pi^2}\int dx\; \left(\partial_x \hat{\theta}_a(x)\right)\left(\partial_x \hat{\theta}_b(x)\right)
     \\ +\sum_{n=1}^3 \frac{\mu_{n}}{a}\int dx\;\cos{\left(n\left(\hat{\f}_{\Up}-\hat{\f}_{\Dn}\right)\right)}.
\end{multline}
By rotating the fields $\hat{\f}_{c/s}(x)=\left(\hat{\f}_{\Up}(x)\pm \hat{\f}_{\Dn}(x)\right)/\sqrt{2}$ and the corresponding dual ones $\hat{\theta}_{c/s}(x)$, we obtain the Hamiltonian \eqref{SG} 
in the main text with the perturbative relations
\be
K_{c/s}=\pi\sqrt{\dfrac{E_J}{\left(2E_c\pm V_\perp\right)}}\qquad \text{and}\qquad u_{c/s}= a\sqrt{E_J\left(2 E_C\pm V_\perp\right)}.
\label{Lpar}
\ee
In general, a finite intra-leg capacitance $C_L$ among adjacent islands leads to a long range interaction stemming from the inverse capacitance matrix \cite{Fazio2001rev} with screening length $\lambda=a\sqrt{C_L/C_g}$, where $C_g$ is the self capacitance. However, this may be ignored as long as one is interested in the physics of modes with energies lower than $u_{c/s}/\lambda$.

From a perturbative point of view the plasma frequency of the spin sector $u_{s}/a=\Lambda \simeq \sqrt{E_J \left(2E_c-V_\perp\right)}$ defines a UV cut-off that allows us to define the dimensionless coupling $\tilde{\mu}_n=\mu_{n}/\Lambda$ in the sine-Gordon Euclidean action,
\be
S\left[\f_s(x,\tau)\right]=\dfrac{1}{2\pi}\int dxd\tau\;K_s\left(\left(\de_{\tau}\f_s\right)^2+\left(\de_x\f_s\right)^2\right)-\sum_{n=1}^3\frac{\tilde{\mu}_n}{a^2}\int dxd\tau\;\cos{\left(\sqrt{2}n\f_s\right)},
\label{three-freq-SG}
\ee
where we have rescaled the imaginary time $\tau\to u_s\tau$.
The operators $\On=\cos{\left(\sqrt{2}n\hat{\f}_s\right)}$ correspond to primaries of the unperturbed free boson $c=1$ theory with scaling dimensions
\begin{equation}
    \Delta_{n}=\dfrac{n^2}{2K_s}.
\end{equation}
Therefore, such operators drive the LL to a massive phase, namely they are relevant, only when $\Delta_n<2$ inferring the lower bound $K_s>9/4$ considered in the main text to make $\mathcal{O}_{n\leq 3}$ relevant. 

\begin{figure}
    \centering
    \includegraphics[width = 0.6\textwidth]{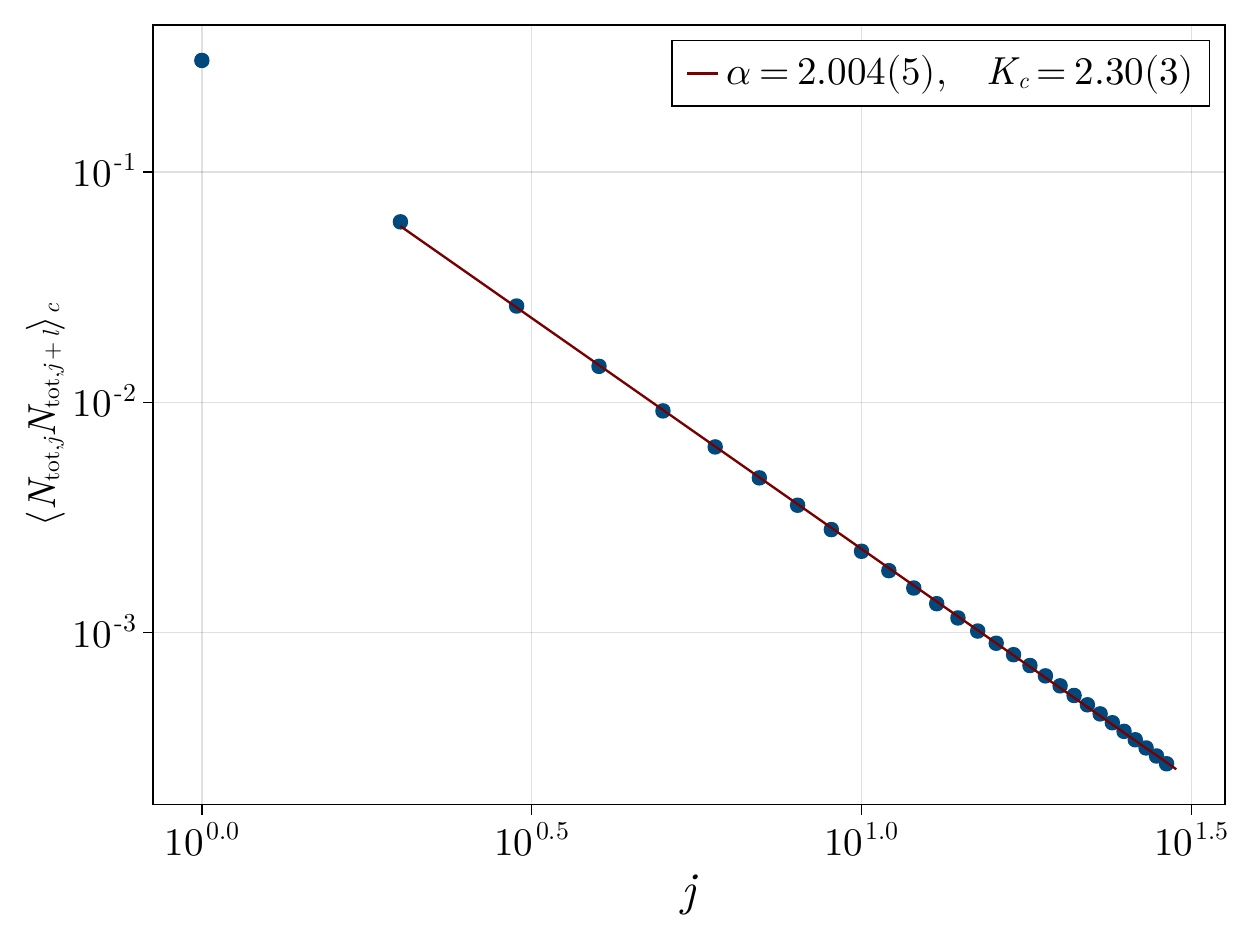}
    \caption{Connected part of the correlation functions of the total density operator $\widehat{N}_{\text{tot},j} = \left(\opN_{\Up,j}+\opN_{\Dn,j}\right)$ taken at
    a random position in the phase-diagram 
    [$X_1 \approx -0.3$ and $X_2 \approx 0.47$] with $T_2=0.6$. The red line is the result of a fit by a function $f(j) = K_c/\pi^2  j^{-\alpha}$.
    The fit result $\alpha \approx 2$ well reproduces the predictions
    from bosonization theory, and also the obtained Luttinger parameter is close to the prediction from perturbation theory: $K_c^{\rm pert}\approx2.61$.}
    \label{SMfig3}
\end{figure}

Note that the charge sector remains massless as there is no sine-Gordon potential for $\hat{\varphi}_c$. We checked the validity of this statement in our lattice simulation. In the LL liquid phase the density correlation functions is expected to show the following power-law decay

\begin{equation}
\label{density}
    \braket{\wh{\rho}_{\rm tot}(x) \wh{\rho}_{\rm tot}(y)} \sim 
    \frac{2}{\pi^2}\left\langle \de_x\theta_{c}(x,\tau)\;\de_y\theta_ c(y,\tau)\right\rangle= \frac{K_c} {\pi^2}\frac{1}{\left|x-y\right|^2}.
\end{equation}
In the ladder model, the operator $\wh{\rho}_{\rm tot} (x)$ corresponds to the
total rung density offset $\opN_{{\rm tot},j} - \braket{\opN_{{\rm tot},j}}$
with $\opN_{\rm tot}=\opN_{\Up,j}+\opN_{\Dn,j}$.
We explicitly checked the decay of Eq.~\eqref{density} for each point of the phase diagram by fitting
a power-law decay [Fig.~\ref{SMfig3}]. The so found $K_c$ parameters are in a good agreement with the perturbative approximations
given by Eq.~\eqref{Lpar}. This confirms the validity of the field theoretical approach in the low energy regime
of the ladder.

On the other hand, the spin sector \eqref{three-freq-SG} is subject to the different relevant interactions in Eq. \eqref{three-freq-SG} which tend to order the SC phase difference $\hat{\f}_s$. In Ref. \cite{Toth2004} the author shows that this quantum field theory flows to a tricritical Ising point with central charge $c=7/10$ for suitable values of the coupling constants $\mu$. Despite the absence of any non-perturbative mappings between our lattice operators and the massless excitations of this field theory, we can exploit the Ginzburg-Landau representation of the TCI CFT to gain insight about this relation.

The operator content of the CFT is split in the odd and even sector with respect to the $\setZ_2$-symmetry and is characterized by 6 primary fields: the identity $I$, four relevant operators $\sigma,\;\epsilon\;\sigma',\;\epsilon'$ ($\Delta<2$) and one irrelevant $(\Delta>2)$ operator. 
The Ginzburg-Landau Lagrangian representation of the TCI corresponds to \cite{mussardobook}
\begin{equation}
    \mathcal{L}=\dfrac{K_s}{2\pi}\varphi_s\left(\de_x^2+\dfrac{\de_\tau^2}{u_s^2}\right)\varphi_s-\lambda_2:\varphi_s^2:-\lambda_4:\varphi_s^4:-\lambda_6:\varphi_s^6:,
\end{equation}
where $::$ indicates the normal ordering with respect to the tricritical point CFT. In the mean-field limit $K_s\gg 1$, we can build an approximate mapping bewteen local operators in our theory and the primary fields (see also Ref. \cite{essler16}), 
\begin{equation}
\label{opmap}
    \begin{aligned}
        \varphi_s(x)&\to\sigma(x), \quad \left(h_\sigma,\bar{h}_\sigma\right)=\left(\dfrac{3}{80},\dfrac{3}{80}\right)\\
        :\varphi_s^2(x):&\to\epsilon(x), \quad \left(h_\epsilon,\bar{h}_\epsilon\right)=\left(\dfrac{1}{10},\dfrac{1}{10}\right)\\
         :\varphi_s^3(x):&\to\sigma'(x), \quad \left(h_{\sigma'},\bar{h}_{\sigma'}\right)=\left(\dfrac{7}{16},\dfrac{7}{16}\right)\\
         :\varphi_s^4(x):&\to\epsilon'(x), \quad \left(h_{\epsilon'},\bar{h}_{\epsilon'}\right)=\left(\dfrac{3}{5},\dfrac{3}{5}\right),\\
    \end{aligned}
\end{equation}
which implies the expansion of the local order operator $\hat{J}_\perp$ in terms of the most relevant operator $\sigma$ close to the critical point,
\begin{equation}
    \hat{J}_\perp(x) = \sin{\left(\sqrt{2}\hat{\f}_s(x)\right)\sim \hat{\f}_s(x)+\ldots\to\sigma(x)+\ldots}
\end{equation}
In the previous expansion the dots indicate less relevant operator contributions. 

\section{Charge basis}

For the numerical simulations, we formulated the Hamiltonian \eqref{Ham} 
from the main text in the charge basis. In this basis the operator $\wh{N}_{\spec, \pos}$
is diagonal and defines how the number of Cooper pairs differs from the
average occupation on the island $(\spec, \pos)$:
\be
    \wh{N}_{\spec, \pos} = \mathrm{diag}\left(\dots,-2, -1, 0, 1, 2, \dots \right)\, .
\ee
Using this choice, it is easy to show that $e^{i\hat{\varphi}_{\spec,\pos}}$ must to be of the form
\be
   e^{i\hat{\varphi}_{\spec,\pos}}=\begin{pmatrix}
    &\ddots&& & && &\\
    &&0&1&&&&\\
    & & & 0&1& &&\\
    & & &  &0&1 & &\\
    &&&&&  \ddots&&\\
    \end{pmatrix}_{\spec,\pos}\equiv\wh{\Sigma}_{\spec,\pos}^-
\ee
for the commutator $[\wh{N}, \wh{\Sigma}^-] = -\wh{\Sigma}^-$ to hold.
Further, in order to represent these operators in our simulations, we have to truncate the number of possible charge states
\be
    \wh{N}_{\spec, \pos} = \mathrm{diag}\left(-N_{\rm max}\dots,-2, -1, 0, 1, 2, \dots  N_{\rm max}\right)\, ,
\ee
i.e. we adopt a truncated local Hilbert-space of dimension $2N_{\rm max} + 1$ per each SC island.
We can control the error caused by this truncation by varying $N_{\rm max}$ until we reach convergence in all observables.
Alternatively, we can measure the probability $\braket{\hat{P}^n_{\spec,\pos}}$ of finding an excitation $n$ on the island $(\spec, \pos)$.
By ensuring that $N_{\rm max}$ is large enough to have negligible weight 
$\braket{\hat{P}^{N_{\rm max}}_{\spec,\pos}} < \epsilon$ we can claim to be converged in $N_{\rm max}$.
In practice we found that $N_{\rm max} = 8$ gives $\braket{\hat{P}^{N_{\rm max}}_{\spec,\pos}} \sim 10^{-9}$.
The Hamiltonian used for the simulation finally reads $\wh{H} = \sum\limits_{\pos=0}^L \wh{h}_{\pos, \pos+1}$ with:
\be
    \begin{split}
    \hat{h}_{\pos, \pos+1} =&     
    \sum_{\alpha = a,b} \left[
        E_c\left(\wh{N}_{\spec,\pos} \right)^2 
        - \frac{E_J}{2} \left( 
            \wh{\Sigma}_{\spec, \pos}^+  \wh{\Sigma}_{\spec, \pos + 1}^- + 
            \wh{\Sigma}_{\spec, \pos}^-  \wh{\Sigma}_{\spec, \pos + 1}^+ 
        \right)
        \right]\\
        {}&+ V \wh{N}_{a, \pos} \wh{N}_{b, \pos}
        + \frac{\mu_{1}}{2} \left(
        \wh{\Sigma}_{a, \pos}^+  \wh{\Sigma}_{b, \pos}^- +
        \wh{\Sigma}_{b, \pos}^+  \wh{\Sigma}_{a, \pos}^-
        \right) \\
        {}&+ \frac{\mu_{2}}{2} \left(
        \left(\wh{\Sigma}_{a, \pos}^+\right)^2  \left(\wh{\Sigma}_{b, \pos}^-\right)^2 +
        \left(\wh{\Sigma}_{b, \pos}^+\right)^2  \left(\wh{\Sigma}_{a, \pos}^-\right)^2
        \right) \\
        {}&+ \frac{\mu_{3}}{2} \left(
        \left(\wh{\Sigma}_{a, \pos}^+\right)^3  \left(\wh{\Sigma}_{b, \pos}^-\right)^3 +
        \left(\wh{\Sigma}_{b, \pos}^+\right)^3  \left(\wh{\Sigma}_{a, \pos}^-\right)^3
        \right) 
    \end{split}
\ee

\section{Further numerical evidence for the transitions} \label{app:transfer}

In this section, we present additional numerical indications about the different nature of the transitions across the phase diagram. All the data in this section refer to a system with $T_2=0.6$, but variations of the parameter $T_2$ do not affect qualitatively our results as long as $T_2$ is sufficiently large to observe the symmetry-broken phase.

\subsection{Hysteresis and gap jump at the first-order transition}

First of all, we present additional evidence of first-order phase transitions (FOPTs) along the horizontal cuts at $X_2=0.52$ (between the disordered phase I and the ordered phase II) and at $X_2=0.6$ (between phases I and III).

\begin{figure}[t]
    \centering
    \includegraphics[width = \textwidth]{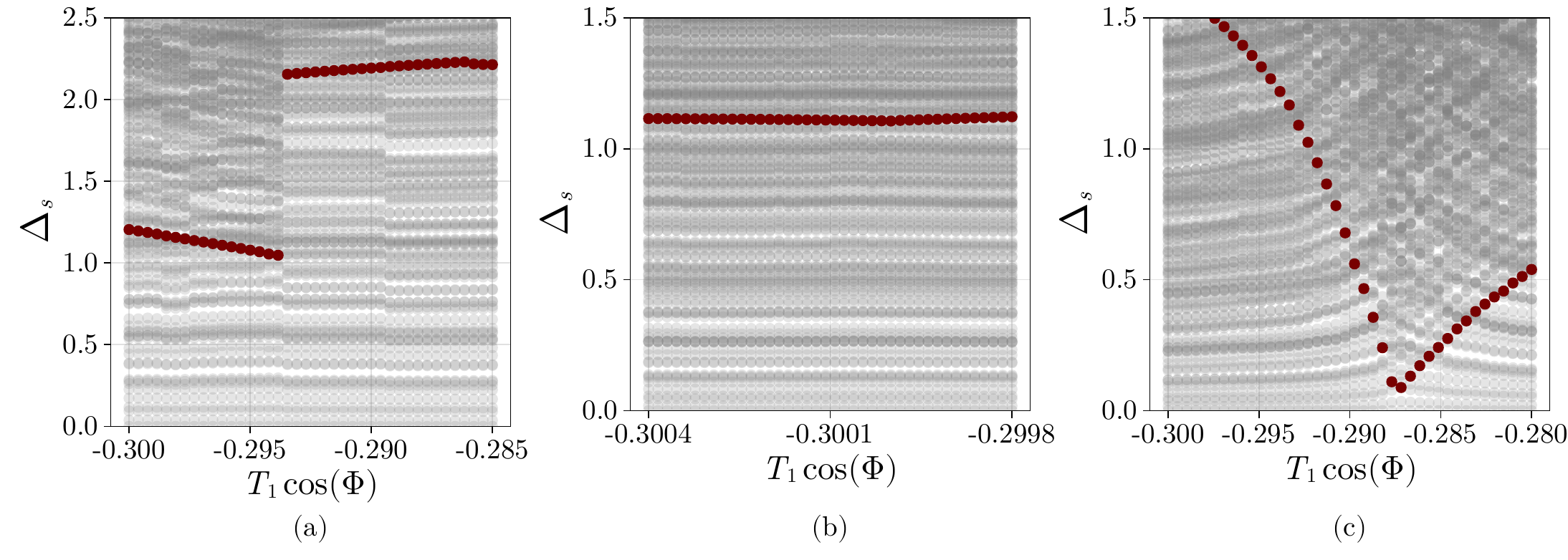}
    \caption{The gap in the spin sector $\Delta_s$ is determined by tracking the second largest eigenvalue of the transfer matrix within the spin sector $\lambda_1$. The results are obtained for the three cuts shown in the main text: a) $X_2 = 0.6$, b) $X_2 = 0.52$ and
    c) $X_2 = 0$. The red points correspond to the gap $\Delta_s=-\log(\lambda_1)$ which remains finite across the FOPTs in panel (a) and (b), while displaying the gap-closing feature of a second-order phase transition in panel (c).}
    \label{SMfig4}
\end{figure}

One significant indicator involves the distinct behavior of the lowest energy excitation in the spin sector. Its energy corresponds to the system's gap, which can be extracted (see Section \ref{secS5}) from the transfer matrix spectrum as shown in Fig.~\ref{SMfig4}. By following the corresponding eigenvalue of the transfer matrix $\lambda_1$, we can extract the gap of the spin sector $\Delta_s=-\log{\lambda_1}$. Across a second-order phase transition, the physical gap closes and, in the numerical VUMPS simulations, this is marked by a minimum in $\Delta_s$ [panel (c)] which approaches zero by increasing the bond dimension. Across a FOPT, instead, the spin gap remains finite [panels (a) and (b)], although it may display a discontinuity when the mass of the spin excitations is different in the two phases. Panels (a) and (b) respectively depict the typical behaviors of the FOPT between the two disordered phases and between phase II and phase I. In the latter case, the related order parameter displays a very weak variation, resulting in an almost continuous behavior of $\Delta_s$.

\begin{figure}[ht]
    \centering
    \includegraphics[width = \textwidth]{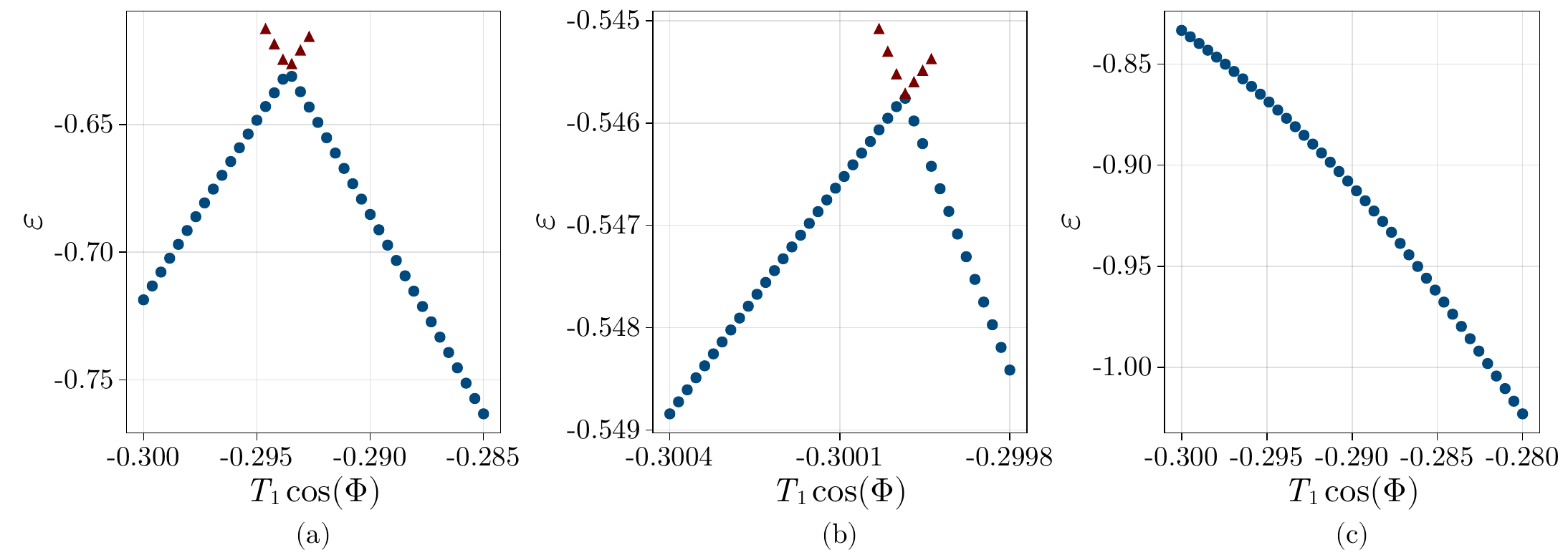}
    
    \caption{
    Energy density $\varepsilon$ of the groundstate obtained at
    the three cuts from the main text: a) $X_2 = 0.6$, b) $X_2 = 0.52$ and
    c) $X_2 = 0$. The red triangles in the case of a) and b) are obtained
    by minimizing the Hamiltonian $H({\bf X} + \delta)$ by starting
    from one of the two groundstates left/right of the meeting point of the
    two branches. The minimization procedure follows these branches
    instead of falling into the true ground state.
    The absence of such an effect for $X_2=0$ is another indication
    for a FOPT in the case a) and b), but a second order phase transition for 
    c).
    }
    \label{SMfig5}
\end{figure}

This behavior is reflected also in the analysis of the hysteresis in the order parameter and the many-body ground state energy, as illustrated in Fig.~\ref{SMfig5}.

A discontinuity in the first derivative of the energy density is observed in the FOPT cases, which is absent in the second-order transition at $X_2=0$ and indicates the crossing of the lowest energy levels. Furthermore, by altering the minimization procedure at each point $X_1$ and initializing the ground state with the result from $X_1\pm\delta$, the variational algorithm follows the corresponding branch, even within the opposite phase. This can be interpreted as a hysteresis effect induced by the orthogonality of these two states around the crossing point.

Also in this case the features of the FOPT are stronger between the two disordered phases -- panel~\ref{SMfig5} (b) is depicted with a magnified energy scales with respect to panel (a). The discontinuity of the derivative $\partial \varepsilon / \partial X_1$ is around $30\;E_J$ in panel (a) and $22\;E_J$ in panel (b). This is physically related to the jump of the average loop current circulating around each triple JJs element, namely $\hat{J}_{\rm loop}=\de \hat{H}/\de \Phi$.

\subsection{Scaling and critical exponents Ising phase transition}

\begin{figure}[h]
    \centering
    \includegraphics[width = 0.9\textwidth]{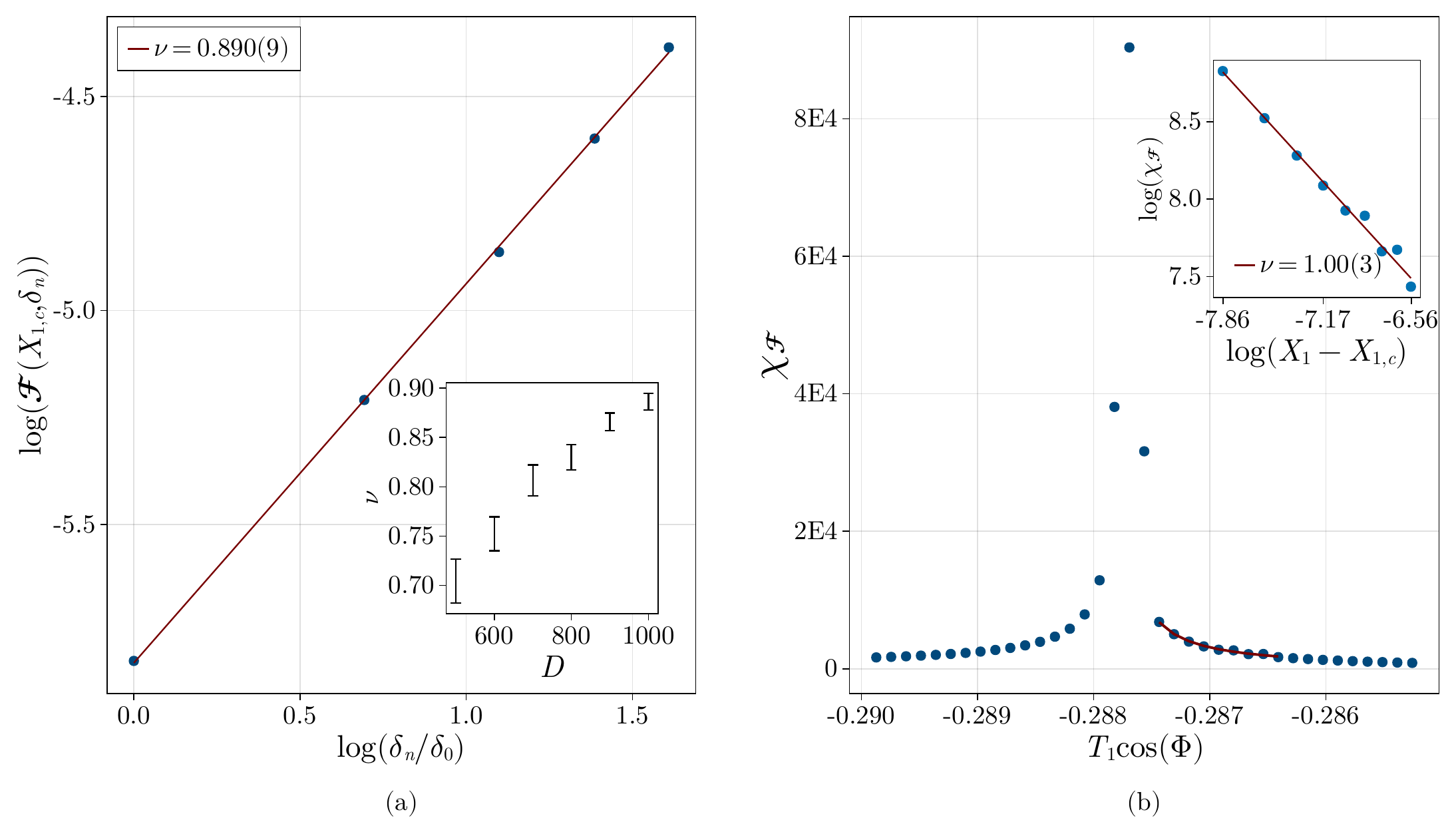}
    \caption{
    Extrapolation of the critical exponent $\nu$ from the scaling features of the fidelity at $X_2=0$. (a): Two-parameter fit of the relation $\mathcal{F}\left(X_{1c},\delta\right)= a\left|\delta\right|^{\nu}$ at bond dimension $D=1000$. The extracted values of $\nu$ increase with the increasing bond dimension (inset). (b): Fit of the fidelity susceptibility $\chi_\mathcal{F}=b \left|X_1-X_{1c}\right|^{\nu-2}$ away from the critical point with a fixed $\delta\ll \left|X_1-X_{1c}\right|$. The plot in log-log scale is shown in the inset. The position of the critical point $X_{1c}$ is obtained from the collapse of the spin correlation length \eqref{collapse}.}
    \label{SMfig6}
\end{figure}

In this subsection, we focus on characterizing the critical exponents $\nu$ and $\beta$, which describe how the correlation length diverges and the order parameter approaches zero across the continuous phase transitions. Concerning the Ising line, we will consider as main example the $X_2=T_1\sin(\Phi)=0$ cut corresponding to Fig.~\ref{fig3}(c)-(d) 
of the main text. In this case, the measured values indicate indeed that the transition belongs to the Ising universality class with $\nu_{\rm IS}=1$ and $\beta_{\rm IS}=1/8$.
To extract these exponents, we relied on scaling properties of three different quantities: the log-fidelity per site $\mathcal{F}$ (and its susceptibility $\chi_\mathcal{F}$), the correlation length of the spin sector $\xi_s$ and the order parameter $\hat{J}^{(2e)}_\perp$.

We determine the critical exponent $\nu$ through two different methods based on the fidelity scaling, both yielding values near $\nu_{\rm IS}=1$ [Fig.~\ref{SMfig6}]. The first approach involves fitting the non-analytic behavior of the log-fidelity per site at the critical point, showing a consistent increase towards $\nu=1$ as the bond dimension $D$ grows [Fig.~\ref{SMfig6}(a), inset], although the adopted bond dimensions were not sufficient to converge to $\nu=1$. The second approach, instead, provides more accurate results and relies on analyzing the divergence pattern of the fidelity susceptibility along a horizontal cut; in this way we obtain $\nu=1.00(3)$ [Fig.~\ref{SMfig6}(b)].

To take into account finite bond dimension corrections, we employed the finite entanglement scaling discussed in Ref. \cite{Tagliacozzo2008} for the spin correlation length $\xi_s$. 
Similarly to finite size effects, the finite bond dimension introduces an artificial length scale making all
correlation functions exponential decaying even at critical points.
This can be interpreted as the addition of a relevant perturbation of the underlying CFT.
However, in the $D \to \infty$ limit, the gapless nature of the model must be restored. 
This artificial length scale is associated with the critical exponent $\kappa$:
\[
\xi_D \sim D^\kappa
\]
and we use this relation to define the following scaling ansatz \cite{Tagliacozzo2008}
\begin{equation}
     \xi_D = D^\kappa f\left(D^{\frac{\kappa}{\nu}} \frac{|X_1 - X_{1c}|}{X_{1c}} \right) \,,\quad
     f(x) \sim \begin{cases}
         \text{const}\ ,\,&x \to 0\\
         \frac{1}{x^\nu}\ ,\,&x\gg 1
        \end{cases}
    \label{collapse}
\end{equation}
where $\nu$ is the critical exponent of the correlation length in the infinite bond dimension case. We use this ansatz and the collapse procedure explained in \cite{Bhattacharjee2001} to determine the critical point $X_{1c}$ and to extract the critical exponents $\nu$ and $\kappa$ discussed in the main text.

\begin{figure}[h]
    \centering
    \includegraphics[width =0.6 \textwidth]{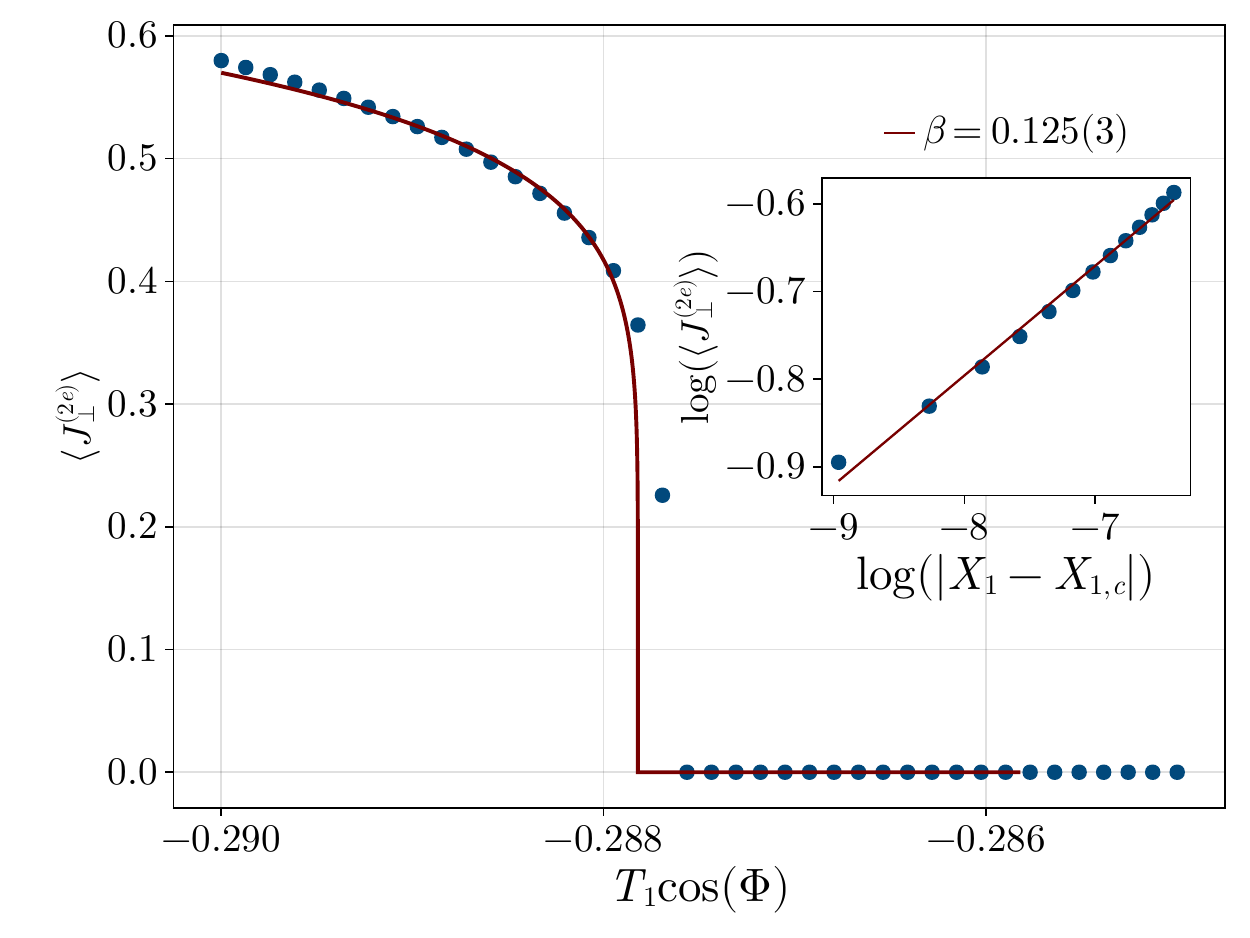}
    \caption{The single-particle current $J^{(2e)}_\perp$ plays the role of the effective magnetization at the Ising critical point, displaying a scaling behavior $|X_1-X_{1c}|^\beta$ with the fitted value $\beta=0.125(3)$ (red curve). The critical point $X_{1c}$ is fixed by the collapse of $\xi_s$ obtained by using Eq. \eqref{collapse}. The discrepancy with the numerical points is due to finite entanglement effects that shifts the position of the critical point at finite bond dimensions.}
    \label{SMfig7}
\end{figure}

 Additionally, to extract the critical exponent $\beta$ we employ the scaling of the expectation value of the single-particle current $\hat{J}^{(2e)}_\perp$ close to the critical point. Indeed, this operator plays the role of the Ising magnetization which is odd under the $\setZ_2$-symmetry $\hat{\f}_s\to-\hat{\f}_s$. By fitting the expected scaling behaviour $\left|X_1-X_{1c}\right|^\beta$, we obtain the critical exponent $\beta=0.125(3)$ [Fig.~\ref{SMfig7}] at $X_2=0$, and analogous values are obtained for $|X_2|\lesssim 0.435$, as depicted in Fig.~\ref{fig3}(e) in the main text.

These results collectively indicate that our findings concerning the transition from the ordered to the disordered phase sufficiently far from the first order discontinuities are compatible with the Ising universality class with $\nu_{\rm IS}=1$ and $\beta_{\rm IS}=1/8$.

The critical exponents $\kappa$ extracted for the spin correlation length at the second order transitions are typically
smaller than one. This implies that a considerable increase of the bond dimension is required in order to faithfully capture the algebraic decay of correlation functions over a long distance. Taking the example of the $X_2 = 0$ cut from the main text with $\kappa \approx 0.8$. The largest correlation length obtained 
for $X_2$ is $\xi_s \approx 30$ for a bond dimension of $D=1000$. Using the scaling behavior $\xi_s \sim D^{0.8}$ we estimate that
a bond dimension $D^\star \approx 4500$ is necessary to get $\xi_s \approx 100$ sites, and  $D^\star \approx 18000$ for $\xi_s \approx 300$ sites.

\subsection{Central charge}

\begin{figure}[htb]
    \centering
    \includegraphics[width =1 \textwidth]{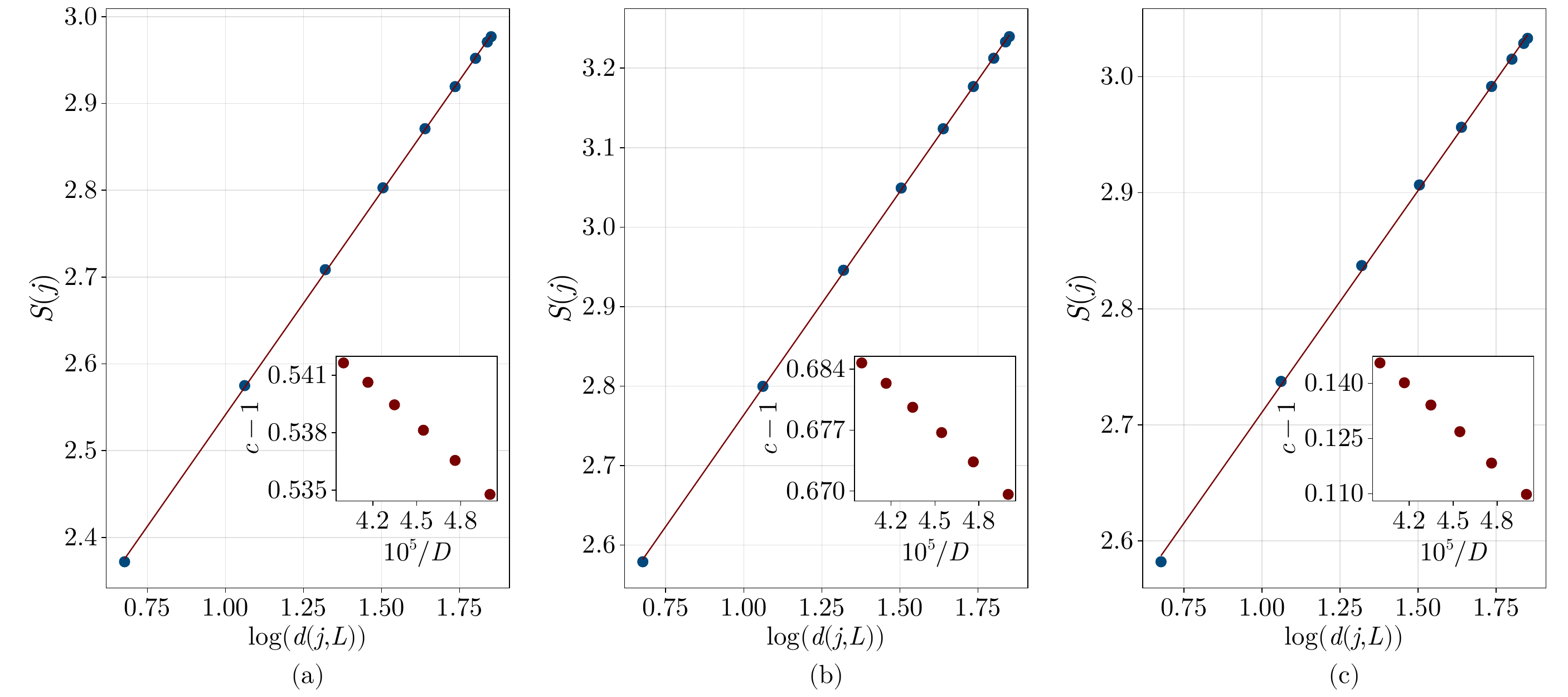}
    \caption{Fits of the entanglement entropy relation \eqref{calabrese-cardy} for $L=20$ and bond dimension $D=2500$ at three significant points along the transition line. The inset shows the slow convergence of
    the fitted value with respect to the inverse of the bond dimension, allowing for an extrapolation $D\to\infty$.
    For $X_{2c}\simeq 0.42$ (a), this interpolation yields $c\approx0.57$. At $X_{2c}\simeq 0.464  $ (b), the central charge increases, $c\approx0.74$ before dropping for $X_{2c}\simeq 0.479 $ (c).}
    \label{SMfig8}
\end{figure}

Given the separation of the two sectors in our model, in the thermodynamic limit the entanglement entropy of the system is predicted to display a typical divergence $S=c_c/6\log(\xi_c) + c_s/6\log(\xi_s)$ \cite{Calabrese2004} in proximity of the second-order phase transition, with $c_{c/s}$ the central charge of the charge/spin sector.
However, strong finite entanglement effects in the VUMPS simulations have a quantitative impact on the estimate of the latter and result in strong fluctuations. Moreover, the theory of finite-entanglement corrections \cite{Pollmann2009,Tagliacozzo2008,Rams2011} is less developed than the finite-size scaling and, in particular, doesn't cover the case of two gapless modes sharing the same finite bond dimension in the MPS representation. In particular, as already pointed out at the end of previous section, achieving a reliable description of the critical correlations of the system with $\xi_s\to\infty$ requires a very large bond dimension $D$, given the sub-linear scaling of $\xi_s\sim D^\kappa$.

For these reasons, we determined the total central charge $c$ from finite-size DMRG simulations with periodic boundary conditions by fitting the relation  \cite{Calabrese2004}
\begin{equation}
\label{calabrese-cardy}
S(j)=\frac{c}{3}\log\left(d\left(j,L\right)\right)+s_1,
\end{equation}
where $S(j)$ is the entanglement entropy at the site $j$, $d(j,L)=L/\pi \sin\left(\pi j/L\right) $ is the chord distance, and $s_1$ is a non-universal constant.

We specifically traced the transition line where the VUMPS spin correlation length $\xi_s$ is maximal and the critical exponent $\beta$ shows the CFTs predictions before vanishing at the FOPT, \ref{fig3}(e) 
in the main text.
Figure \ref{SMfig8} shows the excellent agreement of our data with the relation \eqref{calabrese-cardy} at three illustrative points along this line. Finite size effects are present in any case and lead to an overestimation of the value of the central charge. The measured estimate is expected to decrease by increasing the size of the finite system.

\section{Extraction Of Correlation Lenghts}\label{secS5}
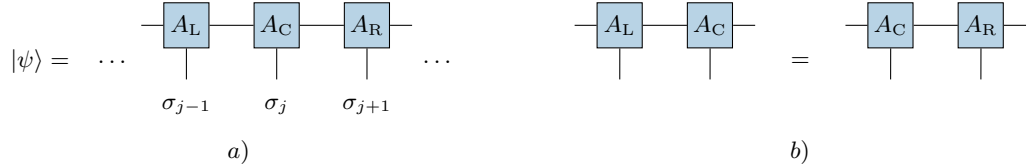
\begin{figure}[ht]
    \centering
    \begin{tikzpicture}[baseline, scale = 1.2]
        \node[anchor = east] at (0.0,0) {$\ket{\psi} = $};
        
        \node at (0.4, 0) {$\dots$};
        \node[aten] (a1) at (1.2, 0.4) {$A_{\rm L}$};
        \node[aten] (a2) at (2.2, 0.4) {$A_{\rm C}$};
        \node[aten] (a3) at (3.2, 0.4) {$A_{\rm R}$};

        \draw (a1) -- ($(a1) - (0.5, 0)$);
        \draw (a3) -- ($(a3) + (0.5, 0)$);

        \node at ($(a1) - (0, 0.9)$) {$\sigma_{j-1}$};
        \node at ($(a2) - (0, 0.9)$) {$\sigma_{j}$};
        \node at ($(a3) - (0, 0.9)$) {$\sigma_{j+1}$};
        
        \draw (a1) -- (a2);
        \draw (a2) -- (a3);
        \draw (a1) -- ($(a1) - (0,0.6)$);
        \draw (a2) -- ($(a2) - (0,0.6)$);
        \draw (a3) -- ($(a3) - (0,0.6)$);
        
        \node at (4, 0) {$\dots$};

        \node[aten] (al) at (6, 0.4) {$A_{\rm L}$};
        \node[aten] (ac) at (7, 0.4) {$A_{\rm C}$};

        \node at (8, 0) {$=$};

        \node[aten] (ac2) at (9, 0.4) {$A_{\rm C}$};
        \node[aten] (ar) at (10, 0.4) {$A_{\rm R}$};

        \draw (al) -- (ac);
        \draw (ac2) -- (ar);

        \draw (al) -- ($(al) - (0.5,0)$);
        \draw (ac) -- ($(ac) + (0.5,0)$);
        \draw (ar) -- ($(ar) + (0.5,0)$);
        \draw (ac2) -- ($(ac2) - (0.5,0)$);
        
        \draw (al) -- ($(al) - (0,0.6)$);
        \draw (ac) -- ($(ac) - (0,0.6)$);
        \draw (ac2) -- ($(ac2) - (0,0.6)$);
        \draw (ar) -- ($(ar) - (0,0.6)$);
        
        \node[anchor=east] at (2,-1) {$a)$};
        \node at (8, -1) {$b)$};
        
    \end{tikzpicture}
    \caption{(a): VUMPS ansatz in the central gauge. 
    (b): Equivariance property to ensure translational invariance}\label{spin_corr:fig:gauged_vumps}
\end{figure}

\begin{figure}[ht]
    \centering
    \begin{tikzpicture}[scale = 1.2]
        \node[anchor=east] at (0,0) {$\braket{\hat{O}_{j}\hat{K}_{j +l}}\coloneqq$};

        \node at (0.4, 0) {$\dots$};

         \foreach \j in {1,...,3}{
            \node[aten] (Al\j) at (\j, 0.75) {$A_{\rm L}$};
            \node[aten] (Ald\j) at (\j, -0.75) {$\bar{A}_{\rm L}$};
        }
        \node[oins] (O) at (2, 0) {$O$};
        
        \draw (Al1) -- (Al2);
        \draw (Al2) -- (Al3);
        \draw (Ald1) -- (Ald2);
        \draw (Ald2) -- (Ald3);
        \draw (Al1) -- (Ald1);
        \draw (Al2) -- (O);
        \draw (O) -- (Ald2);
        \draw (Al3) -- (Ald3);

        \draw ($(Al1) - (0.6, 0)$) -- (Al1);
        \draw ($(Al3) + (0.6, 0)$) -- (Al3);
        \draw ($(Ald1) - (0.6, 0)$) -- (Ald1);
        \draw ($(Ald3) + (0.6, 0)$) -- (Ald3);

        \node at (4, 0) {$\dots$};
        \node[aten] (Al4) at (5, 0.75) {$A_{\rm L}$};
        \node[aten] (Ald4) at (5, -0.75) {$\bar{A}_{\rm L}$};
        \draw (Al4) -- (Ald4);
        \draw ($(Al4)  - (0.6, 0)$) -- (Al4);
        \draw ($(Ald4) - (0.6, 0)$) -- (Ald4);

        \node[aten] (Ac) at (6, 0.75) {$A_{\rm C}$};
        \node[aten] (Acd) at (6, -0.75) {$\bar{A}_{\rm C}$};
        \node[oins] (K) at (6, 0) {$K$};
        \draw (Ac) -- (K);
        \draw (K) -- (Acd);
        \draw (Al4) -- (Ac);
        \draw (Ald4) -- (Acd);
        \node[aten] (Ar) at (7, 0.75) {$A_{\rm R}$};
        \node[aten] (Ard) at (7, -0.75) {$\bar{A}_{\rm R}$};

        \draw ($(Ar) + (0.6, 0)$) -- (Ar);
        \draw ($(Ard) + (0.6, 0)$) -- (Ard);
        \draw (Ar) -- (Ard);
        \draw (Ac) -- (Ar);
        \draw (Acd) -- (Ard);

        \node at (7.6, 0) {$\dots$};

        \draw [decorate, thick,
                decoration = {calligraphic brace, 
                              amplitude=5pt, 
                              raise = 12pt}] (3,0.75) -- (5, 0.75)
                              node[pos=0.5,above=20pt,black]{$l-1$};;

        \node[anchor=east] at (8.2,0) {$=$};
        \node[aten] (Al1) at (9, 0.75) {$A_{\rm L}$};
        \node[oins] (O) at (9, 0) {$O$};
        \node[aten] (Ald1) at (9, -0.75) {$\bar{A}_{\rm L}$};
        \draw (Al1) -- (O);
        \draw (O) -- (Ald1);
        
        \draw (Al1) edge[bend right=90] (Ald1);

        \node [aten] (Ac) at (12, 0.75) {$A_{\rm C}$};
        \node [aten] (Acd) at (12, -0.75) {$\bar{A}_{\rm C}$};
        \node[oins] (K) at (12, 0) {$K$};

        \draw (Ac) edge[bend left=90] (Acd);
        \draw (Ac) -- (K);
        \draw (Acd) -- (K);
        
        \node[inner sep = 0] (Tu) at (10.5, 0.75) {};
        \node[inner sep = 0] (Td) at (10.5, -0.75) {};

        \draw (Al1) -- (Tu);
        \draw (Ald1) -- (Td);
        \draw (Ac) -- (Tu);
        \draw (Acd) -- (Td);
        
        \node[tmat, fit=(Al1)(Ald1), inner sep = 0] (T) at (10.5, 0) {$\mathcal{T}_{\rm L}$};

        \draw (10.9,-1.25) to [ncbar=0.5cm,out=60,in=-60] (10.9, 1.25);
        \draw(10.1, -1.25) to [ncbar=0.5cm,out=120,in=-120] (10.1, 1.25);
        \node at (11.4, 1.4) {$l-1$};
        
    \end{tikzpicture}
    \caption{Correlation function in the infinite system.}
    \label{spin_corr:fig:transfer_mat_decomp}
\end{figure}
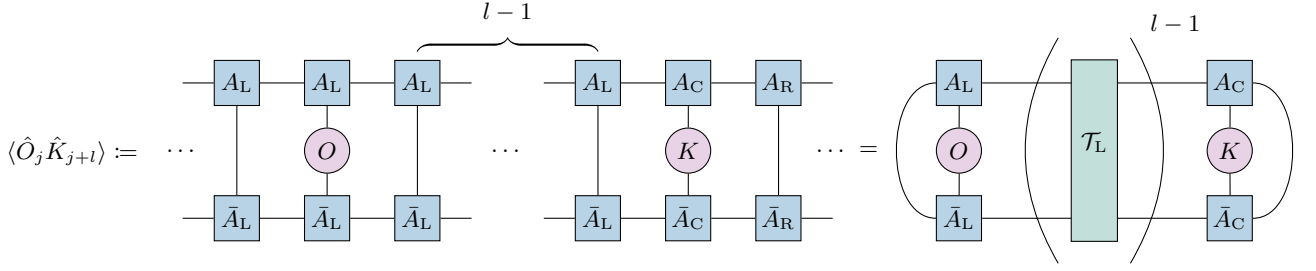

Most of the numerical results presented in this latter are obtained by the VUMPS algorithm presented in Ref. \cite{Stauber2018}.
The concrete implementation uses the \rm{ITensor} library \cite{Fishmann2022}.
This ansatz operates directly in the thermodynamic limit by enforcing translational invarance. 
The class of ansatz states is characterized by the set of matrices $\lbrace A_L^\sigma, A_C^\sigma, A_R^\sigma\rbrace$, with $\sigma$ enumerating the physical local states. From this set of matrices, the state $\ket{\psi}$ is represented as
\[
    \ket{\psi} = \sum_{\lbrace \sigma\rbrace} \Tr\left[ 
        \dots A_L^{\sigma_{j-2}} A_L^{\sigma_{j-1}} A_C^{\sigma_{j}} A_R^{\sigma_{j+1}} A_L^{\sigma_{j+2}} \dots
    \right] \ket{\dots \sigma_{j-2} \sigma_{j-1}\sigma_{j}\sigma_{j+1}\sigma_{j+2} \dots} \, .
\]
The matrices $A_L^\sigma$ and $A_R^\sigma$ fulfill $\sum_\sigma (A_L^\sigma)^\dagger A_L^\sigma = \sum_\sigma A_R^\sigma (A_R^\sigma)^\dagger = \unit$ and
special equivariance relations to ensure the translational invariance of the ansatz, see Fig.~\ref{spin_corr:fig:gauged_vumps}.
Using the transfer-matrix of the system, defined by
\begin{equation}
    \label{spin_corr:eq:transfermatrix}
     \mathcal{T}_{L} \coloneqq \sum_{\sigma} A_{L}^{\sigma} \otimes \bar{A}_{L}^{\sigma}\,,
\end{equation}
and the two transfer-matrices with operator insertion
\begin{equation}
    \mathcal{T}_L^{O} \coloneqq \sum_{\sigma, \tau} O_{\sigma, \tau}
    A^{\sigma}_L\otimes\bar{A}^\tau_L\,,\quad
    \mathcal{T}_C^{K} \coloneqq \sum_{\sigma, \tau} K_{\sigma, \tau}
    A^{\sigma}_C\otimes\bar{A}^\tau_C\,,\quad
\end{equation}
where $\bar{z}$ denotes the complex conjugation of $z$,
one can represent the correlation function of two arbitrary operators $\hat{O}$ and $\hat{K}$ as, Fig.~\ref{spin_corr:fig:transfer_mat_decomp}:
\begin{equation}
    \label{spin_corr:eq:corr_fn}
    \begin{split}
        \braket{\hat{O}_{j}\hat{K}_{j +l}}  &= \bra{\unit}\mathcal{T}^O_{\rm L}
        \left(\mathcal{T}_{\rm L}\right)^{l-1}
        \mathcal{T}^K_{\rm C}\ket{\unit} = \sum_{n\ge0} \lambda_n^{l-1} \alpha^O_n\, \beta^K_n 
        = \sum_{n\ge0} e^{-\frac{l-1}{\xi_n}}c^{O,K}_n\\
        \alpha^O_n &= \braket{\unit | \mathcal{T}_O | R_n},\ 
        \beta^K_n = \braket{L_n | \mathcal{T}_K | \unit},\ \xi_n = -\frac{1}{\log(\lambda_n)} \, .
    \end{split}
\end{equation}
The second line in Eq.~\ref{spin_corr:eq:corr_fn} is obtained after using the eigen decomposition of the transfer-matrix
\begin{equation}
    \label{spin_corr:eq:eigendecomp}
    \mathcal{T}_L = \sum_{n\ge 0} \lambda_n \ket{R_n}\bra{L_n}
    \,,\quad
    \braket{L_n|R_m} = \delta_{m,n}\,.
\end{equation}
Using Eq.~\ref{spin_corr:eq:corr_fn}, it is straightforward to extract the
asymptotic behavior of any correlation function
\[
\braket{\hat{O}_j^\vdag \hat{K}_{j+l}^\dag} 
\approx c_{n^\star}^{O,K}\,e^{\frac{l}{\xi_{n^\star}}} + c_{0}^{O,K} \, .
\]
where $n^\star$ is the first $n>0$ in the descending sequence
$\lambda_0 > |\lambda_1| \ge |\lambda_2| \dots$
with a non-zero operator weight $c_n^{O,K}$ (assuming $\lambda_{n^\star}$ to be unique).
The contribution $c_0^{O,K}$ equals the product of expectation values
$\braket{\hat{O}_j} \braket{K^\dag_j}$.
In the case of $\hat{O} = \hat{K}$ this asymptotic behavior can be used to extract
the smallest energy gap in the excitation spectrum generated by the operator $\hat{O}$.
In the main text, we applied this analysis to the current operator 
\[
\hat{O} = \wh{J}^{(2e)}_\perp \coloneqq 
\frac{i}{2}\left( \Sigma_{\Up}^+ \Sigma_{\Dn}^-
        - \Sigma_{\Dn}^+ \Sigma_{\Up}^- \right)\,.
\]
which can be interpreted as the magnetization order parameter in the field theory $\sin{\left(\sqrt{2}\hat{\f}_s(x)\right)}$
odd under the $\varphi_s(x) \to -\varphi_s(x)$ symmetry transformation. Thus, $\hat{J}^{(2e)}_\perp$
is naturally associated to excitations in the spin-sector exclusively.

\bigskip

Very similarly, one can extract the density of the logarithmic fidelity $\mathcal{F}$ in the thermodynamic limit from the mixed transfer-matrix
\begin{equation}
     \mathcal{T}_L^{\phi,\psi} \coloneqq \sum_{\sigma} A_{L}^{\phi,\sigma} \otimes \bar{A}_{L}^{\psi, \sigma}\, ,
\end{equation}
where $A_L^\phi$ defines the state $\ket{\phi}$ and $A_L^\psi$ the state $\ket{\psi}$. Define $\lambda_0$ the smallest in maginutde eigenvalue of
$\mathcal{T}_L^{\phi,\psi}$, it is straigthforward to show:
\[
    \mathcal{F} \coloneqq -\lim_{N \to \infty} \frac{1}{N}\log\left(\braket{\psi|\phi}\right) = -\log(|\lambda_0|) \, .
\]

\twocolumngrid

\bibliography{TCI_bibliography}

\end{document}